\documentclass[aps,pre,twocolumn,floats]{revtex4} 
\usepackage{epsfig} 
\begin{document}
 
\newcommand{\tbox}[1]{\mbox{\tiny #1}} 
\newcommand{\half}{\mbox{\small $\frac{1}{2}$}} 
\newcommand{\mbf}[1]{{\mathbf #1}} 

 
\title{
Driven chaotic mesoscopic systems, dissipation and decoherence \\
{\small (Lecture notes of the course in the 2002 Karpacz school)}
}

\author{Doron Cohen}

\date{Jan 2002, March 2002}
 
\affiliation{
Department of Physics, Ben-Gurion University, Beer-Sheva, Israel.
} 
 
 
\begin{abstract}
Driven quantum systems, described by Hamiltonian
${\cal H}(Q,P,x(t))$ where $x(t)$ is a time dependent parameter,
are of interest in mesoscopic physics (quantum dots),
as well as in nuclear, atomic and molecular physics.
Such systems tend to absorb energy. This irreversible
effect is known as dissipation.
More generally, $x$ may be a dynamical variable,
where the total Hamiltonian is ${\cal H}_0(x,p)+{\cal H}(Q,P;x)$.
In such case the interaction of $(x,p)$ with the
environmental degrees of freedom $(Q,P)$ leads to
dephasing as well as to dissipation.
It should be emphasized that even few $(Q,P)$ degrees
of freedom can serve as a miniature heat bath,
provided they have chaotic dynamics.
We shall introduce a general framework for the analysis
of dissipation and dephasing, and we shall clarify the
tight connection to recent studies of quantum
irreversibility (also referred to as "Loschmidt echo"
or as the "fidelity" of quantum computation).
Specific model systems that will be presented are:
particle in a box driven by moving a wall,
and particle in a box/ring driven by electro-motive-force.
These two examples are related to studies of nuclear
friction and mesoscopic conductance.
Specific issues to be discussed are the limitations
of kinetic theory, the capabilities of linear response
theory, and the manifestation of non-perturbative
quantum-mechanical effects. In particular we shall
explain that random matrix theory and the semiclassical
theory lead to different non-perturbative limits.
\end{abstract}


\maketitle

\section{Introduction}

In the following lectures we are interested in
systems that are described by a Hamiltonian
${\cal H}(Q,P;x(t))$, where $(Q,P)$ is a set
of (few) canonical coordinates, and $x(t)$
is a time dependent parameter. We further assume
that in the time independent case ($x(t)=\mbox{const}$)
the classical motion is chaotic.

The Quantum mechanical (QM) study of classically
chaotic systems is known in the literature
as "quantum chaos". The theory of time independent
Hamiltonians, in particular their spectral properties,
is well documented \cite{gutz,haake,sto}.
But the QM theory of driven chaotic systems
is still a missing chapter.

To avoid misunderstanding we refer here to systems
that are chaotic also in the absence of driving.
By "driving" we mean that we assume
Hamiltonian ${\cal H}(Q,P;x(t))$ where $x(t)$
is time dependent, rather than constant.
Unlike the case of time independent systems,
in case of driven systems the energy distribution
evolves with time. What one needs is a theory for
this evolution. We shall see that various notions,
such as "dissipation", "irreversibility" and "fidelity"
just emphasize particular aspects of this evolution.

The available theory for driven systems in the
quantum mechanical literature is based mainly upon
first order perturbation theory,
supplemented by higher order (sometimes non-perturbative)
corrections. Depending on the "rate" of the driving,
one commonly distinguishes between the "adiabatic"
and the "Fermi golden rule" regimes \cite{wilk,WA}.
The latter is known also as the
"linear response theory" (LRT)
or as the "Kubo-Greenwood" regime.

There are circumstances where first order perturbation
theory cannot serve as a starting point for the analysis
of a driven system. However the well known examples refer
to systems that are {\em not} chaotic in the absence of driving.
This includes in particular one-degree-of-freedom systems
such as the quantum kicked rotator \cite{qkr}.
Our interest is in generic chaotic systems,
therefore we have to consider systems that
have at least two degrees of freedom.

The possibility to present a general QM theory
for driven chaotic systems \cite{vrn,crs,rsp,frc}
follows from the simple fact that "chaos" leads to universality.
This universality is captured, to some extend,
by random matrix theory (RMT).
On the other hand we have semiclassical methods.
We shall see that there is a clash between RMT and semiclassics,
and that they lead to different non-perturbative limits \cite{wbr,wpk}.


\section{Model Systems}

The classic example for a driven system is
the piston model (Fig.1), where a gas in confined
inside a cylinder, and $x$ is the position
of the piston. Our interest is in the
case where we have "one particle gas".
[Note however that if we know to solve the problem
for one particle, then automatically we can get
the solution for many non-interacting particles].

The 1D-box version of this model (Fig.2a)
is known in the literature as the
"infinite-well" problem with moving wall
\cite{infwell_a,infwell_b}.
Some limited aspects of this problem
have been discussed in the literature in
connection with the Fermi acceleration problem \cite{jose}.

A 2D-box variation of the "piston model" is
presented in Fig.2b. Here we have stadium shaped
billiard, and the the parameter $x$ controls the
deformation of the boundary.
Two other variations of the same model are
presented in Fig.3, where the box has the shape
of a generalized Sinai billiard.

In the examples so far the parameter $x$
controls the shape of the "box", and
$V=\dot{x}$ has the interpretation of
wall velocity. The interest in such systems
has emerged long time ago in studies of nuclear friction
(one-body dissipation) \cite{wall,koon}.
A renewed interest is anticipated
in mesoscopic physics where the shape
of a quantum dots can be controlled
by gate voltages. [Note that in the nuclear
physics context the shape is close to
spherical, while a quantum dot is typically
strongly chaotic].

We can create driving by changing any
parameter (or field). In Fig.4 the driving
is achieved by changing the perpendicular
magnetic field. Fig.4a assumes "quantum dot
geometry" with homogeneous magnetic field,
while Fig.4b assumes Aharonov-Bohm ring geometry
with magnetic flux that goes via the hole.
Let us define $x$ as the total magnetic flux.
In such case $V=\dot{x}$ is the electro-motive
force (measured in Volts) which is induced
in the ring according to Faraday law.

If the variations of the parameter $x$
are classically small, then we can linearize
the Hamiltonian as follows
\begin{eqnarray} \label{e1}
{\cal H}(Q,P;x) = {\cal H}_0(Q,P) + x{\cal W}(Q,P)
\end{eqnarray}
where without loss of generality we have assumed
that $x=0$ is the typical value of $x$.
For generic systems (which means having smooth Hamiltonian
that generates a classically chaotic motion),
the representation of ${\cal W}$,
in the ordered ${\cal H}_0$ determined basis,
is known to be a banded matrix
(for details see the next section).
A simple example can be found in \cite{lds} where
\begin{eqnarray} \label{e2}
{\cal H}(Q,P;x) = \half(P_1^2{+}P_2^2 + Q_1^2{+}Q_2^2)
+ (1+x)\cdot Q_1^2 Q_2^2
\end{eqnarray}
This Hamiltonian describes a particle moving inside
a two dimensional anharmonic well (2DW). The shape
of the 2DW in controlled by the parameter $x$.
The perturbation is ${\cal W}(Q,P)=Q_1^2Q_2^2$,
and its matrix representation $\mbf{B}_{nm}$
is visualized in the inset of Fig.7.

The above discussion of generic Hamiltonian models,
such as the 2DW model, motivates the definition of a
simple artificial model Hamiltonian, that has
the same characteristics: This is Wigner model \cite{wigner,felix1}.
In the following definition of Wigner model we
follow closely the notations of \cite{wbr}.
In the standard representation ${\cal H}_0=\mbf{E}$
is a diagonal matrix whose elements are the ordered
energies $\{E_n\}$, with mean level spacing $\Delta$,
and ${\cal W}=\mbf{B}$ is a random banded matrix
with non-vanishing couplings within the band $0 < |n-m| \le b$.
These coupling elements are zero on the average,
and they are characterized by the variance
$\sigma=(\langle |\mbf{B}_{nm}|^2 \rangle)^{\tbox{1/2}}$.
Hence the Hamiltonian is
\begin{eqnarray} \label{e3}
{\cal H} = \mbf{E} + x\mbf{B}
\end{eqnarray}
This artificial model can serve as a reference case
for testing various theoretical ideas. Moreover,
it has been conjectured that such model captures
some generic features of chaotic systems.
[Note that most of the RMT literature deals with
simplified versions of Wigner model, where the
bandwidth equals to the matrix size].


\section{Quantum Chaos}

The notion of chaos in classical mechanics implies
that few degree of freedom system, such as the
Sinai billiard system, exhibit stochastic-like behavior.
This is in contrast to the out-of-date idea that
stochasticity and irreversibility are the outcomes
of having (infinitely) many degrees of freedom.
Chaos means that the motion (eg Fig.5) has exponential
sensitivity to any perturbation or change in initial
conditions. Another way to characterize a chaotic motion
is by its continuous power spectrum (see Fig.6).
This should be contrasted with integrable motion which is
characterized by a discrete (rather than continuous) set
of frequencies.

For sake of later analysis it is useful to define
the "power spectrum" of the motion specifically
as follows.
Let $(Q(t),P(t))$ be an ergodic trajectory that
is generated by the time independent Hamiltonian
${\cal H}(Q,P;x)$. We can define a fluctuating
quantity ${\cal F}(t)=-d{\cal H}/dx$. In case that
$x$ is the displacements of a wall element (eg Fig.3b),
the fluctuating ${\cal F}(t)$ has the meaning
of "Newtonian force". In case that $x$ is the magnetic
flux (eg Fig.4b), the fluctuating ${\cal F}(t)$ has
the meaning of "electric current".
In case of the 2DW model we get
${\cal F}(t)=-{\cal W}(Q(t),P(t)) = -Q_1(t)^2Q_2(t)^2$.
The correlation function of the fluctuating ${\cal F}(t)$
will be denoted by $C(\tau)$ and the power spectrum
of the fluctuations will be denoted by $\tilde{C}(\omega)$.
The latter is the Fourier transform of the former.
The variance of the fluctuation is $C(0)$, the intensity
of the fluctuations is defined as $\tilde{C}(0)$,
and the correlation time is denoted by $\tau_{cl}$.

It is clear that upon quantization we no longer have
chaos. Still, the question arise what are the fingerprints
of the classical chaos on both the spectral properties
of the system, and also on the structure of the eigenstates.
This problem was the focus of intensive studies during
the last decade \cite{gutz,haake,sto}, and it has important applications
in mesoscopic physics \cite{imry,alh_rev,marcus_rev}.

An important observation of "quantum chaos" studies
is that Quantum Mechanics introduce {\em two} additional
energy scales into the problem (rather than only one).
We can take the 2DW model as a generic example.
After rescaling of the classical parameters of the
model, we are left with one dimensionless parameter
(the dimensionless energy). This parameter controls
the nature of the classical dynamics. Upon quantization
we have {\em two} additional (dimensionless) parameters.
One energy scale is obviously the mean level spacing $\Delta$,
which is proportional to $\hbar^d$.
The other energy scale is $\Delta_b=\hbar/\tau_{cl}$,
where $\tau_{cl}$ is the classical correlation time
that characterizes the (chaotic) dynamics.
If $\hbar$ is small then the two energy scales are
very different ($\Delta \ll \Delta_b$).

The significance of the energy scale $\Delta_b$ is
a central issue in "quantum chaos". It turns out that
the statistical properties of the energy spectrum
are universal on the sub-$\hbar$ scale, and obey the
predictions of RMT. On the other hand,
on large energy scale (compared with $\Delta_b$),
non-universal (system specific) features manifest
themselves \cite{berry}.
These features are the fingerprints
of the underlying classical dynamics.
In the context of ballistic quantum dots,
which are in fact billiard systems,
$\Delta_b$ is also know as the "Thouless energy".

There is another way in which the energy scale $\Delta_b$
manifests itself. Let ${\cal W}(Q,P)$ be some observable,
and consider its matrix representation ${\mbf{B}_{nm}}$ in the
basis which is determined by the (chaotic) Hamiltonian.
An example is presented in Fig.7.
It can be argued \cite{mario} that ${\mbf{B}_{nm}}$ is a banded matrix,
and that the bandwidth is $b=\Delta_b/\Delta$.
This is based on a remarkably robust semiclassical expression
that relates the bandprofile to the classical power spectrum:
\begin{eqnarray} \label{e4}
\left\langle\left|\mbf{B}_{nm}
\right|^2\right\rangle
\ \ \approx \ \
\frac{\Delta}{2\pi\hbar} \
\tilde{C}\left(\frac{E_n{-}E_m}{\hbar}\right)
\end{eqnarray}
We can apply this semiclassical relation to the case
where ${\cal W}$ is the "perturbation" as defined
in Eq.(\ref{e1}). This leads to the interpretation
of $\Delta_b$ as the largest "distance" in energy space
that can be realized in a first-order transition.
We can also use the semiclassical relation in reverse,
in order to find/define the classical correlation function
that corresponds to a quantum-mechanical matrix Hamiltonian.
In case of the standard Wigner model we get
\mbox{$C(\tau)=b\sigma^2 \mbox{sin}(\tau/\tau_{cl})/(\tau/\tau_{cl})$},
with the correlation time $\tau_{cl}=\hbar/(b\Delta)$.

\section{Parametric Evolution}

A more recent development
was to consider a parametric set of Hamiltonians,
namely ${\cal H}(Q,P;x)$ where $x$
is a parameter as in the examples of Section~2.
For each value of $x$ we can diagonalize
the Hamiltonian, leading to set of (ordered) eigen-energies
$E_n(x)$, as in the schematic illustration of Fig.8.
The corresponding eigenstates will be denoted by
$|n(x)\rangle$. Their parametric evolution can
be characterized by the parametric kernel
\begin{eqnarray} \label{e5}
P(n|m) = \left|\langle n(x)| m(x_0) \rangle \right|^2
\end{eqnarray}
We shall use the notation $P(r)=P(n-m)=P(n|m)$,
with implicit average over the reference state $m$.
We shall refer to $P(r)$ as the "average spreading profile".
This is in fact, up to scaling, the LDOS (local density
of states, also known as strength function).

Let us characterize the perturbation by the
quantity $\delta x=x-x_0$. The interesting question
is how $P(r)$ evolves as we increase the
perturbation $\delta x$. For the Wigner model the answer
is known long ago \cite{wigner,felix1,alh_brm,rmt_rev}.
$P(r)$ has a standard perturbative structure for very small
$\delta x$.  For larger $\delta x$ it becomes a chopped Lorentzian,
and for even larger $\delta x$ it becomes a semicircle.
We shall denote the border between the standard perturbative
regime and the Wigner regime by $\delta x_c$,
and the border between the Wigner regime and the
non perturbative (semicircle) regime will be denoted
by $\delta x_{\tbox{prt}}$. The explicit expressions are:
\begin{eqnarray} \label{e6}
\delta x_c \ = \
&  \frac{\Delta}{\sigma}
& \ \ \propto \ \ \hbar^{(1{+}d)/2}
\\
\label{e7}
\delta x_{\tbox{prt}} \ = \
& \sqrt{b} \frac{\Delta}{\sigma}
& \ \ = \ \ \frac{2\pi\hbar}
{\tau_{\tbox{cl}}\sqrt{C(0)}}
\end{eqnarray}
where $d$ is the number of freedoms ($d=2$ for billiards).
In order to determine the $\hbar$ dependence we
have used the semiclassical relation Eq.(\ref{e4}),
and the proportionality $\Delta \propto \hbar^d$.
Note that the latter relation, known as Wyle law,
is significant for the determination of $\delta x_c$.
In contrast to that $\delta x_{\tbox{prt}} \propto \hbar$
is in fact independent of $\Delta$.

The generalization of Wigner scenario has been
the subject of our recent research \cite{wls,lds,prm}.
In the general case the standard perturbative structure
evolves into a "core-tail structure",
while for large $\delta x$  it becomes purely non-perturbative.
In the standard perturbative regime ($\delta x \ll \delta x_c$)
most of the probability $P(r)$ is concentrated
in one level ($n=m$). In the extended perturbative regime
most of the probability is concentrated within
a "core" whose width is typically
$\Gamma(\delta x) \sim (\delta x/\delta x_c)^2$.
The "core" is the non-perturbative component which
arise due to non-perturbative mixing of nearby levels.
The "tail" is the outer perturbative component
which is created by first order transitions.

The extended perturbative regime is defined by the requirement
of having separation of energy scales
$\Gamma(\delta x) \ll \Delta_b$. This condition
is trivially satisfied in the "standard perturbative
regime" where $\Gamma\sim\Delta$.
The condition $\Gamma(\delta x) \ll \Delta_b$
is violated in the non-perturbative regime
($\delta x \gg \delta x_{\tbox{prt}}$),
which in fact leads to the determination
of $\delta x_{\tbox{prt}}$ as in Eq.(\ref{e7}).
The theory for $P(r)$ in the non-perturbative regime
is not complete yet. However, it can be argued \cite{wls}
that if $\delta x$ is large enough, then $P(r)$ becomes
of semiclassical nature \cite{felix2}. The case of billiards
with shape deformation requires special considerations
and is of particular interest \cite{wls,prm}.

It is important to realize that the border of the
standard perturbative regime ($\delta x_c$) is related
to the energy scale $\Delta$,
while the border of the extended perturbative
regime ($\delta x_{\tbox{prt}}$), which leads to the
identification of the non-perturbative regime,
is related to the bandwidth $\Delta_b$.

\section{Temporal evolution}

After considering the parametric evolution,
the next logical stage is to consider the actual
(temporal) evolution which is generated by the
time dependent Hamiltonian ${\cal H}(Q,P,x(t))$.
Then, in complete analogy, we can ask
how the energy scales $\Delta$ and $\Delta_b$
are reflected in the actual evolution.
We postpone the discussion of the latter issue
to Section~11.

The purpose of the present and next sections is
to define what does it mean "driving",
and how do we quantify the temporal evolution.
Without loss of generality we assume $x(0)=0$.
We would like to consider the following driving schemes:
\begin{itemize}
\setlength{\itemsep}{0cm}
\item Linear driving
\item One pulse driving cycle
\item Periodic driving
\item Driving reversal scenario
\item Time reversal scenario
\end{itemize}
In the next section we define the various schemes,
some of which are also illustrated in Fig.9.
The evolution is characterized by the obvious
generalization of Eq.(\ref{e5}), namely
\begin{eqnarray} \label{e8}
P_t(n|m) = \left|\langle n(x(t))| U | m(x(0)) \rangle \right|^2
\end{eqnarray}
Here $U$ is the evolution operator,
with implicit dependence on the time $t$.
The parametric kernel Eq.(\ref{e5})
can be regarded as corresponding to the "sudden" limit
where $U \approx 1$. As in the parametric
case we can define an average spreading
profile $P_t(r)$, where $r=n-m$.

There are various practical possibilities available
for the characterization of the distribution $P_t(r)$.
It turns out that the major features of this distribution
are captured by the following three measures:
\begin{eqnarray} \label{e9}
{\cal P}(t) & = &  P_t(r=0)
\\ \label{e10}
\Gamma(t) & = & \mbox{50\% probability width}
\\ \label{e11}
\delta E(t) & = & \Delta \times \left(\sum_r r^2 P_t(r)\right)^{1/2}
\end{eqnarray}
The first measure is the survival probability ${\cal P}(t)$.
The second measure $\Gamma(t)$ is the energy width of the
central $r$ region that contains 50\% of the probability.
[For simplicity of presentation we use here a loose
definition of $\Gamma$ as far as prefactors of order
unity are concerned.]
Finally the energy spreading $\delta E(t)$ is
defined as the square-root of the variance.

It is important to realize that the above three measures
give different type of information about the nature
of the energy spreading profile. In particular
the indication for having a core-tail structure is:
\begin{eqnarray} \label{e12}
\Gamma(t) \ \ \ll \ \ \delta E(t) \ \ \ll \ \ \Delta_b
\end{eqnarray}
The core-tail structure (eg chopped Lorentzian)
is characterized by a "tail" component that contains
a vanishingly small probability but still dominates
the variance. [Note that $\Delta_b=\infty$,
as in the case of un-chopped Lorentzian, would
imply $\delta E(t)=\infty$ irrespective of $\Gamma$.]
In contrast to that a typical semiclassical spreading
profile (as well as Wigner's Semicircle) is characterized by
\begin{eqnarray} \label{e13}
\delta E(t) \ \ \sim \ \ \Gamma(t) \ \ \gg \ \ \Delta_b
\end{eqnarray}
In the latter case, in order to avoid confusion,
it is better not to use the notation $\Gamma(t)$.
[The notation $\Gamma$ has been adopted in the common
diagrammatic formulation of perturbation theory.
This formulation is useful in the extended perturbative
regime in order to derive Wigner's chopped Lorentzian.
In the non-perturbative regime this formulation
becomes useless, and therefore the significance
of $\Gamma$ is lost.]

\section{Driving schemes}

Linear driving means $x(t) = Vt$, where
$V$ is a constant. In such a scenario obviously
$x(t) \ne x(0)$. Still it is convenient to
assume that the chaotic nature of the dynamics
is independent of $x$, and that changes in $x$
are not associated with changes in phase space
volume (no conservative work is being done).
This is manifestly the case for the systems
which are illustrated in Fig.3b and Fig.4b.
[Note that the standard Wigner model does not
have $x$ invariance property, and therefore
the analysis of linear driving for the Wigner model
is an ill defined problem. Attempts to overcome this
difficulty lead to certain subtleties \cite{rsp}].

For all the other driving schemes we assume, without
loss of generality, that $x(0)=x(T)=0$, where $T$
is the period of the driving cycle. The simplest
driving scheme is a rectangular pulse $x(0<t<T)=A$,
which is characterized by its amplitude $A$.
Does the study of rectangular pulses
constitute a good bridge for developing a
general theory for driven systems?
The answer is definitely not.
An essential ingredient in the theory of
driven system is the rate $V$ in which the
parameter $x$ is being changed in time.
Therefore, it is important to consider,
for example, a triangular pulse (Fig.9b).
Such pulse is characterized by both
amplitude $A$ and driving rate $V=2A/T$.
More generally one may consider (Fig.9c) a train
of such pulses ($=$ periodic driving).
In particular one may consider the usual sinusoidal
driving $x(t)=A\sin(\Omega t)$ where $\Omega=2\pi/T$.
In the later case we can define the root-mean-square
driving rate as $V=\Omega A/\sqrt{2}$.

In all these cases we ask, in complete analogy
with the parametric case, what is the
evolution of the energy distribution $P_t(r)$.
Now the evolution is with respect to time,
rather than with respect to $\delta x$.
The different scenarios are distinguished by the
choice of $U$. We shall use the notation
$U[x_A]$ in order to denote the evolution
operator that corresponds to driving scheme $x=x_A(t)$.

The case of rectangular pulse is known in the
literature as "wavepacket dynamics" \cite{heller}.
The particle is prepared in an eigenstate
of the ${\cal H}_0$ Hamiltonian, while the evolution
is generated by the perturbed Hamiltonian
${\cal H}={\cal H}(Q,P; x = A)$.
We may consider more complicated schemes of pulses.
For example rectangular pulse $+A$ followed by
another rectangular pulse $-A$. The question here
is whether the second pulse can compensate
the effect of the first pulse.
We call such scheme "driving reversal".
The evolution operator can be written as
$U=U[x(\mbox{\small rev})]U[x]$ where $x=A$ represents
the rectangular pulse, while $x(\mbox{\small rev})=-A$
is the reversed pulse.
The case of triangular pulse can be regarded
as another particular variation of driving reversal.
In the latter case $x$ represents linear driving
with velocity $V$, and $x(\mbox{\small rev})$
is the reversed driving process with velocity $-V$.

The case of "driving reversal" should be distinguished
from "time reversal" scheme. The latter notion
is explained in the next section.

\section{Two notions of irreversibility}

There are two distinct notions of irreversibility
in statistical and quantum mechanics. One is based
on the "piston model" paradigm (PMP),
while the other \cite{peres} is based on the
"ice cube in cup of hot water" paradigm (ICP).
The recent interest \cite{jalabert,jacq,tomsovic,fdl,prosen,rvs}
in "quantum irreversibility" is motivated by its relevance
to quantum computing.

In the PMP case we say that a process is reversible
if it is possible to "undo" it by driving reversal.
Consider a gas inside a cylinder with a piston (Fig.1).
Let us shift the piston inside. Due to the compression
the gas is heated up. Can we undo the "heating" simply
by shifting the piston outside, back to its original
position? If the answer is yes, as in the case
of strictly adiabatic process, then we say that
the process is reversible.

In the ICP case we consider the melting process
of an ice cube. Let us assume that after some time
we reverse the velocities of all the molecules.
If the external conditions are kept strictly the same,
we expect the ice-cube to re-emerge out of the water.
In practice the external conditions (fields) are
not exactly the same, and as a result we have
what looks like irreversibility.

The mathematical object that should be considered
in order to study PMP is just $P_t(r)$ for a driving
scheme that involves "driving reversal". Namely,
as discussed in the previous section, the evolution
operator is
\begin{eqnarray} \label{e14}
U \ \ = \ \ U[x(\mbox{\small rev})] \ U[x]
\end{eqnarray}
The mathematical object that should be considered in
order to study ICP is again $P_t(r)$, but with driving scheme
that involves "time reversal". Namely, the evolution operator
is defined as
\begin{eqnarray} \label{e15}
U \ \ = \ \ U[x_B]^{-1} \ U[x_A]
\end{eqnarray}
In the latter case, if the external conditions
are in full control ($x_B=x_A$), then we
have complete reversibility ($U=1$).

It is also important to define precisely what is
the {\em measure} for quantum irreversibility.
This is related to the various possibilities
which are available for the characterization
of the distribution $P_t(r)$.
The prevailing possibility is to take the
survival probability ${\cal P}(t)$ as a measure \cite{peres}.
Another possibility is to take the energy
spreading $\delta E(t)$ as a measure.
The latter definition goes well with the PMP,
and it has a well defined classical limit.
Irreversibility in this latter sense implies
diffusion in energy space, which is the reason
for having energy absorption (dissipation) in
driven mesoscopic systems (see section~9).

\section{Wavepacket dynamics, survival probability and fidelity}

Driving schemes with rectangular pulses are the simplest
for both analytical and numerical studies. It is easiest to
consider the survival probability in case of a single
rectangular pulse. The survival amplitude is defined as
\begin{eqnarray} \label{e16}
F(t) & = & \langle \Psi_0 | U[A] | \Psi_0 \rangle \\
\nonumber \ & = &
\left|\left\langle \Psi_0 \Big|
\exp\left( -\frac{i}{\hbar}{\cal H}t \right)
\Big| \Psi_0 \right\rangle\right|
\end{eqnarray}
The survival probability is ${\cal P}(t)=|F(t)|^2$,
in consistency with the definition of Eq.(\ref{e9}).
$F(t)$ is manifestly a Fourier transform of the LDOS,
and therefore we can immediately draw a
conclusion \cite{wls} that the nature of the dynamics
is different depending on the parametric regime
to which the amplitude $A$ belongs.
If $P(r)$ have a core-tail Lorentzian
structure, then we get an exponential decay
${\cal P}(t)=\exp(-\Gamma t)$. On the other hand
if $P(r)$ has a semiclassical structure, then
the decay of ${\cal P}(t)$ is non-universal (system specific).

A similar picture arise in recent studies of
the survival probability for "time reversal"
driving scheme. Here one defines the fidelity
amplitude as
\begin{eqnarray} \label{e17}
F(t) = \langle \Psi_0 | U[A]^{-1} U[0] | \Psi_0 \rangle
\end{eqnarray}
The fidelity, also known as Loschmidt echo,
is defined as ${\cal P}(t)=|F(t)|^2$. The situation
here is more complicated compared with Eq.(\ref{e16})
because we have two LDOS functions \cite{fdl}:
one is the ${\cal H}_0$ weighted LDOS,
and the other is the $\Psi_0$ weighted LDOS.
The two LDOS functions coincide only
if $\Psi_0$ is an eigenstate of ${\cal H}_0$.
In the latter case the $F(t)$ of Eq.(\ref{e17})
reduces (up to phase factor) to Eq.(\ref{e16}).
It turns out that in case of Eq.(\ref{e17})
there is no simple Fourier Transform
relation between $F(t)$ and the LDOS functions.
However, the picture "in large" is the same
as in the case of Eq.(\ref{e16}) \cite{fdl}.
Namely, one has to distinguish between three regimes
of behavior: In the standard perturbative regime
($A<\delta x_c$) one typically encounters a Gaussian
decay \cite{peres}; In the Wigner regime (also called FGR regime)
one typically finds Exponential decay \cite{jacq};
And in the non-perturbative regime one observes a semiclassical
perturbation-independent "Lyapunov decay" \cite{jalabert}.

The study of the survival probability, as described above,
is only one limited aspect of the temporal evolution.
The more general object that should be considered
is $P_t(r)$ as defined in section~5. The major features
of this time evolution are captured by the three measures
that we have defined in Eq.(\ref{e9})-(\ref{e11}).
In Fig.10 we display numerical simulations of
wavepacket dynamics for the 2DW model.
The energy spreading $\delta E(t)$ is plotted as
a function of time. The first panel is
the classical simulation, which in fact coincides
with the "linear response" calculation.
The input for the LRT calculation is $C(\tau)$,
and the result is proportional to the amplitude $A$.
Namely,
\begin{eqnarray} \label{e18}
\delta E(t) = A \times \sqrt{2(C(0)-C(t))}
\end{eqnarray}
In the second panel we display the results of the
quantum mechanical simulations.  For smaller $\hbar$ values
the agreement with the classical LRT calculation is better.
Finally, in the third panel we repeat the quantum
mechanical simulations with a sign randomized Hamiltonian.
This means that we take Eq.(\ref{e3}), and we randomize
the sign of the off-diagonal terms.
The bandprofile, and hence $\tilde{C}(\omega)$ are
not affected by this procedure, which implies that
the LRT calculation gives exactly the same
result. But now we see that as $\hbar$ becomes smaller
the correspondence with the classical result
becomes worse.
Specifically: In (a) and (b) we see a crossover from
ballistic spreading ($\delta E  \propto t$)
to saturation ($\delta E \sim \mbox{const}$)
as implied by Eq.(\ref{e18}).
Only one time scale ($\tau_{\tbox{cl}} \sim 1$) is involved.
In (c), in contrast to that, we see that as $\hbar\rightarrow 0$
an intermediate stage of diffusion ($\delta E  \propto \sqrt{t}$) develops.

How can we explain the above results. Obviously we see that
for small $\hbar$ we cannot trust LRT. What in fact happens is
that we have a crossover from the perturbative
regime ($A <\delta x_{\tbox{prt}}(\hbar)$) to the non-perturbative
regime ($A > \delta x_{\tbox{prt}}(\hbar)$). In the latter case
we get either semiclassical behavior, or RMT behavior.
In other words, random matrix theory and the semiclassical
theory lead to {\em different} non-perturbative limits.
In the semiclassical case the crossover from LRT behavior
to non-perturbative behavior cannot be detected by looking
on $\delta E(t)$. Still the crossover can be detected
by looking on $\Gamma(t)$. See \cite{wpk} for details.

\section{Diffusion in energy space and Dissipation}

In the following sections we discuss the case of
either linear or periodic driving. In such case
the long time behavior of the system is characterized
by diffusion in energy space. Associated with this
diffusion is a systematic increase of the average energy.
This irreversible process of energy absorption is
known as "dissipation".

There is a satisfactory classical theory for dissipation \cite{cl_dsp}.
Some of the mathematical details are subtle,
but the overall physical picture is quite
simple. Without loss of generality the main idea
can be explained by referring to the billiard example
of Fig.1a.  The particle executes chaotic motion,
and we may say that each collision has roughly equal
probability to be with either the inward-going or
with the outward-going wall. As a result the particle
either gain or loose kinetic energy. Thus, the dynamics
in energy space is like random walk, and it can be
described by a diffusion equation. Thus we see that due
to the chaos we have stochastic-like energy spreading.

This classical diffusion process is irreversible
in the PMP sense. Let us assume that we start with
a microcanonical distribution that has definite energy.
If, after some time, we reverse the velocity of the walls,
then obviously we do not get back the initial
microcanonical distribution.

The effect of dissipation is related to the
irreversible stochastic-like diffusion in energy
space. If the diffusion rate were the same irrespective
of the energy, then obviously the average energy
would be constant. But this is not the case.
The diffusion is stronger as we go up in energy,
and as a result the diffusion process is biased.
Thus the average energy systematically grows with
time, and one can derive a general {\em diffusion
dissipation relation} \cite{flc}:
\begin{eqnarray} \label{e19}
\frac{d}{dt}\langle {\cal H} \rangle
\ = \ - \int_0^{\infty} dE \ g(E) \ D_{\tbox{E}}
\ \frac{\partial}{\partial E}
\left(\frac{\rho(E)}{g(E)}\right)
\end{eqnarray}
where $g(E)$ is the density of states,
and $\rho(E)$ is the probability distribution
(eg microcanonical, canonical or Fermi occupation).
The diffusion picture is generally valid in the classical case,
and it is typically valid also in the quantum mechanical case.
[The issue of dynamical localization due to
strictly periodic driving \cite{qkr} is important for
driven 1D system, but not so important in the
case of driven chaotic systems \cite{rsp}.]

There is a simple linear response (Kubo) expression,
that relates the diffusion coefficient to the power
spectrum $\tilde{C}(\omega)$ of the fluctuations:
\begin{eqnarray} \label{e20}
D_{\tbox{E}} \ \ = \ \
\frac{1}{2} \tilde{C}(\Omega) \times V^2
\end{eqnarray}
The diffusion law for short times is
$\delta E(t) = \sqrt{2D_{\tbox{E}}t}$.
This expression is completely analogous
to Eq.(\ref{e18}). In both cases the
spreading is proportional to the amplitude $A$.
[Recall that for periodic driving we define
$V=\Omega A/\sqrt{2}$. In the special case
of linear driving the spreading is proportional to $V$.]
Moreover, as in the case of wavepacket dynamics,
the LRT result is the {\em same} classically and
quantum-mechanically. But again, as in the case
of the wavepacket dynamics, the validity regime
of LRT in the quantum mechanical case is
much smaller (see section.11).

If we combine the above Kubo expression
with the diffusion-dissipation relation we get
\begin{eqnarray} \label{e21}
\frac{d}{dt}\langle {\cal H} \rangle  =
\mu(\Omega) \times V^2
\end{eqnarray}
where $\mu$ is related to the power spectrum
of the fluctuations. Thus we get a
fluctuations-dissipation relation \cite{flc}.
The standard "thermal" fluctuation-dissipation
relation $\mu(0)=\tilde{C}(0)/(2k_BT)$
is obtained from Eq.(\ref{e19}) in case of
canonical $\rho(E)$.

Standard textbook formulations \cite{flc}
takes linear response theory together with thermal
statistical assumptions as a package deal.
Our presentation provides a  more powerful picture.
On the one hand we can discuss non-equilibrium
situation using LRT combined with an appropriate
version of the diffusion-dissipation relation.
On the other hand, we may consider situation
where LRT does not apply. In such case we may get
some (non-perturbative) result for the diffusion,
and later use the diffusion-dissipation relation
in order to calculate the dissipation rate.

\section{Beyond kinetic theory}

The coefficient $\mu$ in Eq.(\ref{e21}) is called
the "dissipation coefficient". In the case where
$x$ is the displacement of a wall element, it is
also known as "friction coefficient",
and in the case where $x$ is a magnetic flux it can
be called "conductance".

Having dissipation rate proportional to $V^2$
is known as "ohmic" behavior. In case of
"friction" it is just equivalent to saying that
there is a friction force proportional to the
velocity $V$, against which the wall is
doing mechanical work.  This mechanical work
is "dissipated" and the gas is "heated up"
in a rate proportional to $V^2$.

In case of "conductance" we may say
that there is a drift current proportional to
the voltage $V$. This is in fact "Ohm law".
The dissipated energy can either be accumulated
by the electrons (as kinetic energy), or it may
be  eventually transfered to the lattice
vibrations (phonons). In the latter case we say
that the ring is "heated up". The rate of the
heating goes like $V^2$ which is in fact "Joule law".

The traditional approach to calculate $\mu$ is
to use a kinetic picture (Boltzmann) which is based
on statistical assumptions. This leads in
case of friction to the "wall formula" \cite{wall,koon}:
\begin{eqnarray} \label{e22}
\mu(\Omega) =
\frac{N}{\mathsf{V}_{\tbox{box}}}
mv_{\tbox{E}} \mathsf{A}_{\tbox{walls}}
\end{eqnarray}
where $N$ is the number of gas particles
(let us say $N=1$), and $v_{\tbox{E}}=\sqrt{2E/m}$.
We also use the notations ${\mathsf{V}_{\tbox{box}}}$
for the volume of the box, and $\mathsf{A}_{\tbox{walls}}$
for the effective area of the moving walls.
In the latter we absorb some geometric factors \cite{frc}.
Application of the traditional kinetic (Boltzmann) approach
in case of conductance leads to "Drude formula":
\begin{eqnarray} \label{e23}
\mu(\Omega) \ \sim \ \frac{N}{{\mathsf A}_{\tbox{dot}}}
\left(\frac{e^2}{m}\tau_{\tbox{col}}\right)
\frac{1}{1+(\tau_{\tbox{col}} \Omega)^2}
\end{eqnarray}
where ${\mathsf A}_{\tbox{dot}}$ is the
area of the "quantum dot",
and $\tau_{\tbox{col}}$ is the average
time between collisions with the walls.

Using Linear response theory (Kubo formula),
as described in the previous section, we can
go beyond the over-simplified picture of
kinetic theory. That means to go beyond Boltzmann picture.
Below we explain under what assumptions we
get the "traditional" kinetic expressions,
and what in fact can go wrong with these assumptions.

The interest in friction calculation has started
in studies of "one body dissipation" in nuclear
physics \cite{wall,koon}.
The "wall formula" assumes that the collisions
are totally uncorrelated. In such case the
power spectrum $\tilde{C}(\omega)$ of ${\cal F}(t)$
is like that of white noise (namely "flat").
By inspection of Fig.6 we can see that this
assumption is apparently reasonable in the limit of
very strong chaos. But it is definitely
a bad approximation in case of weak chaos.
The dynamics of  chaotic system is typically
characterized by some dominant frequencies.
Therefore we have relatively strong response whenever
the driving frequency matches a "natural" frequency
of the system. This can be regarded as
a classical (broad) resonance.
By inspection of Fig.6 we see that a particular
feature is having such resonance around $\omega=0$.
This type of resonance, due to bouncing behavior,
can be regarded as a "classical diabatic effect" \cite{dia}.

Even if the chaos is very strong, the "white noise"
assumption is not necessarily correct:
In \cite{dil,wlf} we explain that for special
class of deformations (including translations,
rotations and dilations) the low frequency response
is suppressed, irrespective of the chaoticity.
This is illustrated numerically in Fig.11.

In case of Drude formula the fluctuating ${\cal F}(t)$
has the meaning of "electric current", and therefore
the power spectrum $\tilde{C}(\omega)$ is the
Fourier transform of the current-current
(or one may say velocity-velocity) correlation
function. Assuming that the velocity-velocity
correlation function decays exponentially in time,
one obtains the Drude result. A careful analysis
of this assumption, and its relation to the
"white noise approximation" of the "wall formula",
can be found in Section~6 of \cite{wlf}.
Fig.12 displays a numerical example. We clearly
see non-universal deviations from the Drude expression,
which reflect the specific geometry of the "quantum dot".

\section{Non-perturbative response}

In the classical case, assuming idealized system,
the crossover to stochastic energy spreading involves
only one time scale, which is $\tau_{\tbox{cl}}$.
Gaussian-like spreading profile is obtained
only for time $t$ much larger than $\tau_{\tbox{cl}}$.
For short times we can use linear analysis in order
to calculate the spreading profile. However, this
analysis has a breaktime \cite{frc}
that we call $t_{\tbox{frc}}(V)$,
where $V$ is the rate in which $x$ is being changed.
For long times ($t \gg \tau_{\tbox{cl}}$) we can
use stochastic picture. Classical LRT calculation
of the diffusion is valid only if the crossover
to stochastic behavior is captured by the short
time analysis. This leads to the classical
slowness condition
$\tau_{\tbox{cl}} \ll t_{\tbox{frc}}(V)$
which we assume from now on. See specific examples
in Sections 13 and 14.

In the quantum mechanical case we follow a similar
reasoning. The linear analysis is carried out
using perturbation theory. We have presented \cite{frc}
a careful analysis to determine the
breaktime $t_{\tbox{prt}}(V;\hbar)$ of this analysis.
It turns out that this breaktime is not related
to the mean level spacing $\Delta$, but rather to the
much larger energy scale  $\Delta_b$.

In complete analogy with the classical analysis,
it turns out that the validity of LRT in the quantum
domain is restricted by the condition
$\tau_{\tbox{cl}} \ll t_{\tbox{prt}}(V;\hbar)$.
If this inequality is not satisfied, then we say
that we are in a non-perturbative
regime. It is important to realize that the
$\hbar\rightarrow 0$ limit is a non-perturbative limit.
This means that the semiclassical regime is
contained within the non-perturbative regime.

In the simple examples that are discussed in
Sections 13 and 14, the non-perturbative regime
is in fact a semiclassical regime.
This coincidence does not hold in general \cite{vrn,crs,frc}.
In case of RMT models, obviosly we do not
have a semiclassical limit. In such models
the non-perturbative response deviates
significantly from Kubo formula (Fig.13).
The interest in such models can be physically
motivated by considering transport in quantized
{\em disordered} systems.
Whether similar deviations from Kubo formula
can be found in case of quantized {\em chaotic} systems
is still an open question \cite{rsp}.
In any case, it is important to remember
that the rate of dissipation is just one aspect
of the energy spreading process.
Even if Kubo formula does not fail
(thanks to quantum-classical correspondence),
still there are other features of the
dynamics that are affected by the crossover
from the perturbative to the non-perturbative regime.
For example: in Sec.19 we are going to show that
different results are obtained for the dephasing time,
depending whether the process is perturbative
or non-perturbative.

We can express the condition for being
in the non-perturbative regime as \cite{frc}
\begin{eqnarray} \label{e24}
V \ \ \gg \ \ \frac{\delta x_{\tbox{prt}}}{\tau_{cl}}
\end{eqnarray}
The expression in the right hand side
scales like $\hbar$, which reflects
that this condition is related to
$\Delta_b$ and not to $\Delta$.
In the next section we discuss the
definition of the adiabatic
regime (very small $V$) whose existence
is related to having finite $\Delta$.
A schematic illustration of the three
regimes (adiabatic, LRT, non-perturbative)
is presented in Fig.14.
Some further reasoning \cite{rsp} allows
to define the non perturbative regime
in case of periodic driving.
Its location in $(A,\Omega)$ space
is also illustrated in Fig.14.
Note that for periodic driving
we define $V=\Omega A/\sqrt{2}$.
The two $V=\mbox{const}$ curves in the
$(\Omega,A)$ diagram represent the same
conditions as in the case of linear driving.
Other details of this diagram are discussed
in the next section, and in \cite{rsp}.

\section{Adiabatic response and QM resonances}

Let us assume that we are in the perturbative
regime (which means that the non-perturbative regime
of the previous section is excluded).
We ask the following question: can we use the
classical Kubo result as an approximation for
the quantum mechanical result?
The answer is "yes" with the following restrictions:
(i) The amplitude of the driving should be large
enough; (ii) The frequency of the driving should
be large enough. The two conditions are
further discussed below. If they are satisfied
we can trust the classical result. This follows
from the remarkable quantal-classical correspondence
which is expressed by Eq.(\ref{e4}). We have
an illustration of this remarkable correspondence
in Figures 7 and 11.

Large enough amplitude means $A\gg\delta x_c$.
One may say that large-amplitude driving leads
to effective "broadening" of the discrete levels,
and hence one can treat them as if they form
a continuum. This is essential in order to justify
the use of Fermi golden rule (FGR) for a small
isolated system \cite{rsp}. Kubo formula can be regarded as
a consequence of FGR. If the driving amplitude
is not large enough to "mix" levels, we cannot
use FGR, but we can still use first order
perturbation theory as a starting point.
Then we find out, as in atomic physics applications,
that the response of the system is vanishingly small
unless the driving frequency $\omega$ matches
energy level spacing. This is called
"QM resonance". The strips of QM resonances
are illustrated in the schematic diagram of Fig.14.
It is important to realize that higher order of
perturbation theory, and possibly non-perturbative
corrections, are essential in order to calculated
the non-linear response in this regime \cite{wilk}.
Still, first order perturbation theory is
a valid starting point, and therefore we do not
regard this (non-linear) regime as "non-perturbative".

Large enough frequency means $\omega \gg \Delta/\hbar$.
The remarkable quantal-classical correspondence
which is expressed by Eq.(\ref{e4}) is valid only
on energy scales that are large compared with $\Delta$.
If this condition is not satisfied, we have to take
into account the level spacing statistics \cite{robbins,ophir}.
This means that we can have significant difference
between the quantal LRT calculation, and the classical
LRT calculation.

However, this is not the whole story. If $V$ is
small enough, first order perturbation theory
implies "QM adiabaticity". The condition for
QM adiabaticity is $V\ll\delta x_c/t_{\tbox{H}}$
where $t_{\tbox{H}}=2\pi\hbar/\Delta$ is the
Heisenberg time. A useful way of writing the
QM adiabaticity conditions is:
\begin{eqnarray} \label{e25}
V \ \ \ll \ \ \frac{1}{b^{3/2}}
\left(\frac{\delta x_{\tbox{prt}}}{\tau_{cl}}\right)
\end{eqnarray}
In the adiabatic regime, first order perturbation
theory implies zero probability to make a transition
to other levels. Therefore, to the extend that we can
trust the adiabatic approximation, all the probability
remains concentrated in the initial level.
Thus, in this regime, as in the case of small
amplitudes, it is essential to use higher orders
of perturbation theory, and non-perturbative
corrections (Landau-Zener \cite{wilk}).
Still we emphasize that first order perturbation
theory is in fact a valid starting point,
and therefore we do not regard this (non-linear) regime
as "non-perturbative".

\section{Driving by electro-motive force}

Consider a charged particle moving inside
a chaotic ring. Let $x$ represent a magnetic
flux via the ring. If we change $x$ in time,
then by Faraday law $V=\dot{x}$ is the electro-motive force
(measured in Volts). The fluctuating
quantity ${\cal F}(t)$ has the meaning
of electric current. The variance of the
fluctuations is $C(0)=(ev_{\tbox{E}}/L)^2$,
where $v_{\tbox{E}}=\sqrt{2E/m}$,
and $L$ is the length of the ring.
The correlation time of these fluctuations
is the ballistic time
$\tau_{cl}=\tau_{\tbox{col}} = L_{\tbox{col}}/v_{\tbox{E}}$.

Having characterized the fluctuations, we can
determine the bandwidth $\Delta_b=\hbar/\tau_{\tbox{col}}$.
A straightforward calculation leads to the result:
\begin{eqnarray} \label{e26}
b \ \ = \ \
\left[\frac{L}{L_{\tbox{col}}}\right] \times
\left( \frac{L_{\perp}}{\lambda_{\tbox{E}}} \right)^{d{-}1}
\end{eqnarray}
where $\lambda_{\tbox{E}}=2\pi\hbar/(mv_{\tbox{E}})$
is the De-Broglie wavelength,
and $L_{\perp}$ is the width of the ring.
Using Eq.(\ref{e7}) we can determine the
non-perturbative parametric scale:
\begin{eqnarray} \label{e27}
\delta x_{\tbox{prt}} \ \ = \ \
\left[\frac{L}{L_{\tbox{col}}}\right] \times \frac{\hbar}{e}
\end{eqnarray}
which up to a geometric factor equals
the quantal flux unit.
Note that in order to mix levels
a relatively small change in the flux
is needed, as implied by comparing
Eq.(\ref{e6}) with Eq.(\ref{e7}).

We turn now to the analysis of the spreading
in the time dependent case, say for linear driving.
The classical "slowness" condition
which has been mentioned in section~11
is simply $eV \ll E$ where $E$ is the kinetic energy
of the charged particle. Upon quantization
we should distinguish the non-perturbative
regime using Eq.(\ref{e24}), leading to
\begin{eqnarray} \label{e28}
eV \ \ \gg \ \ \left[\frac{L}{L_{\tbox{col}}}\right]
\frac{\hbar}{\tau_{\tbox{col}}}
\end{eqnarray}
Disregarding the geometric prefactor, the quantity
in the right hand side is the so called Thouless energy.
We also should distinguish the QM adiabatic regime
using Eq.(\ref{e25}), leading to
\begin{eqnarray} \label{e29}
eV \ \ \ll \ \
\left(\frac{\lambda_{\tbox{E}}}{L}\right)^{3/2}
\frac{\hbar}{\tau_{\tbox{col}}}
\end{eqnarray}
where we have assumed for simplicity
a simple 2D quantum dot geometry as in Fig.4a.

\section{Driving by moving walls}

There is an ongoing interest \cite{infwell_a,infwell_b}
in the problem of 1D box with moving wall (also known as
the infinite well problem with moving wall).
If the wall is moving with constant velocity,
then it is possible to transform the Schrodinger
equation into a time-independent equation,
and to look for the stationary states.

We are interested in the dynamics, and therefore
we would like to go beyond this limited scope
of study. Before we discuss the general case,
it is useful to point out the $d>1$ generalization
of the above picture.
We can easily show that for any "special deformation"
which is executed in either constant "velocity" or
"acceleration", we can transform the
Schrodinger equation into a time-dependent
equation. By "special deformation" we mean
either translation or rotation or dilation,
or any linear combination of these.
The statement is manifestly trivial for translations
and rotations (it is like going to a different
reference frame), but it is also true
for dilations. The 1D box with moving wall
is just a special case of dilation.

It is important to realize that in case of generic
deformation of chaotic box, we cannot "eliminate"
the time dependence. Thus it is not possible
to reduce the study of "dynamics" to a search
for "stationary solutions".

The determination of $\Delta_b$ for this
system is quite obvious but subtle \cite{prm}.
As one can expect naively the result
is $\Delta_b=2\pi\hbar/\tau_{\tbox{col}}$,
where $\tau_{col}$ is the mean time between
collisions with the moving walls.
The subtlety here is that we cannot
interpret $\Delta_b$ as "bandwidth".
Formally the correlation time of
${\cal F}(t)$ is $\tau_{cl}=\infty$ which implies
infinite bandwidth. Still, some non-trivial
reasoning \cite{wls,prm} leads to the conclusion that the
naive result (rather than the "formal" one)
is in fact effectively correct. A straightforward
calculation leads to the result:
\begin{eqnarray} \label{e30}
b \ \ = \ \
\frac{{\mathsf V}_{\tbox{box}}}
{L_{\tbox{col}} \lambda_{\tbox{E}}^{d{-}1}}
\ \ = \ \
\frac{{\mathsf A}_{\tbox{walls}}}
{\lambda_{\tbox{E}}^{d{-}1}}
\end{eqnarray}
where $\lambda_{\tbox{E}}=2\pi\hbar/(mv_{\tbox{E}})$
is the De-Broglie wavelength, and ${\mathsf V}_{\tbox{box}}$
is the volume of the box, and $L_{\tbox{col}}$
is the mean free path between collisions.
As for the effective value of $\delta x_{\tbox{prt}}$,
again the details are subtle, but the naive
guess turns out to be correct. With the proper
(natural) choice of units for the displacement
parameter $x$, the result is simply
\mbox{$\delta x_{\tbox{prt}}=\lambda_{\tbox{E}}$}.

The way to analyze the dynamics for box with moving
walls is outlined in \cite{wld}.
The classical LRT domain is $V \ll v_{\tbox{E}}$,
where $v_{\tbox{E}} = \sqrt{2E/m}$.
Upon quantization we should distinguish the
non-perturbative regime using Eq.(\ref{e24}),
leading to
\begin{eqnarray} \label{e31}
V  \ \ \gg \ \ \frac{\hbar}{mL_{\tbox{col}}}
\end{eqnarray}
In the non-perturbative regime the dynamics
has a semiclassical nature, and the energy spreading
process has a resonant random-walk nature.
This should be contrasted with the behavior in
the perturbative non-adiabatic regime,
where Fermi-golden-rule (FGR) picture applies.

We also should distinguish the QM adiabatic regime
using Eq.(\ref{e25}), leading to
\begin{eqnarray}
V \ \ \ll \ \
\left(\frac
{\lambda_{\tbox{E}}^{d{-}1}}
{{\mathsf A}_{\tbox{walls}}}
\right)^{3/2}
\frac{\hbar}{mL_{\tbox{col}}}
\end{eqnarray}
In the QM adiabatic regime the spreading is dominated
by transitions between near-neighbor levels:
This is the so called Landau-Zener spreading
mechanism \cite{wilk}. See also Section~20 of \cite{frc},
and the numerically related work in \cite{diego}.


\newpage

\section{Brownian motion}

Brownian motion is a paradigm for the general problem
of system that interacts with its environment.
(See Fig.15 and general discussion in the next section).
One can imagine, in principle, a "zoo" of models that
describe the interaction of a Brownian particle with its
environment. However, following Caldeira and Leggett \cite{zcl},
the guiding philosophy is to consider "ohmic models"
that give Brownian motion that is described by
the standard Langevin equation in the classical limit.
Four families of models are of particular interest:
\begin{itemize}
\setlength{\itemsep}{0cm}
\item Interaction with chaos.
\item Interaction with many-body bath.
\item Interaction with harmonic bath.
\item Interaction with random-matrix bath.
\end{itemize}
Below we assume that the total Hamiltonian
has the following general form
\begin{eqnarray}
{\cal H}_{\tbox{total}} \ = \ {\cal H}_0(x,p) \ + \ {\cal H}(Q,P;x)
\end{eqnarray}
where $(x,p)$ are the system coordinates,
and $(Q,P)$ are the environmental degrees of freedom.

Interaction with chaos provides the simplest
model for Brownian motion \cite{jar}.
See Fig.16a for illustration of the model.
The large Brownian particle is described
by the canonical coordinates $(x,p)$, while
the gas particles are described by the
canonical coordinates $(Q,P)$. It is
important to realize that in order to have
Brownian motion, it is not essential to
consider "many particle gas". "One particle gas"
in enough, but the motion of the gas particle
should be chaotic.

The fluctuations of the environment are
in fact (according to our definition in Section~3)
the random-like collisions of the gas particle
with the Brownian particle. These fluctuations
are like "noise". If the motion of the gas particle
is strongly chaotic, then the power spectrum of these
fluctuations (Fig.5) is just like that of white noise.
[This characterization is meaningful up to
a cutoff frequency which is determined
by the rate of the collisions.]

On the other hand we have the effect of dissipation.
If the particle is launched with a velocity $\dot{x}=V$,
then the rate of dissipation is proportional to $V^2$
as explained in section~9.  Having dissipation implies
that the Brownian particle experiences friction force
which is proportional to the velocity $V$. This is the
reason why the dissipation coefficient is known also
as friction coefficient.

Interaction with chaos can be regarded as
the "mesoscopic" version of Brownian motion.
Our interest in this set of lectures is
in this type of interaction. We want to know
whether {\em few} degree of freedoms can serve
as a "bath". Before we further get into
this discussion we would like to describe
the "conventional" point of view
regarding Brownian motion. The rest of this
section is dedicated for this clarification.

The conventional point of view regarding Brownian motion
assumes an interaction with many body bath. We can consider
a bath that consists of either Bosons or Fermions \cite{zwerger,vacchini}.
The emerging models are quite complicated for analysis,
and therefore, as already mentioned above, it is more common
to adopt a phenomenological approach.

Interaction with (many body) harmonic bath is not
very natural, but yet very popular model
for Brownian motion. In order to have "white noise"
(at high temperatures or in the classical limit)
we should make a special assumption regarding
the frequency distribution of the bath oscillators.
This is known in the literature as the "ohmic choice".
[The characterization of the noise as "white"
is valid up to some cutoff frequency. The latter is determined
by the specific choice of the frequency distribution.]
Also here, as in the case of interaction with chaos,
we have fluctuation-dissipation theorem that implies
"ohmic" dissipation rate (proportional to $V^2$).

There is still some freedom left in the definition of
the interaction with the harmonic bath. This leads to the
introduction of the Diffusion-Localization-Dissipation (DLD)
model \cite{dld,gbm,qbm}. This model gives in the classical limit
Brownian motion which is described by the standard
Langevin equation (white noise + ohmic dissipation).
The familiar Zwanzig-Caldeira-Leggett (ZCL) model \cite{zcl}
can be regarded as a special limit of the DLD model.
The physics of the ZCL and of the DLD model is illustrated
in Fig.16b and Fig.16c respectively, and the model
Hamiltonians can be visualized by the drawings of Fig.17.
The ZCL model describes a motion under the influence of
a fluctuating homogeneous field of force which is induced
by the environmental degrees of freedom. In case of the DLD model
the induced fluctuating field is further characterized
by a finite correlation distance.

For completeness we note that random-matrix modeling of the
environment, in the regime where it has been solved \cite{blgc},
leads to the same results as those obtained for the DLD model.

\section{System interacting with environment}

The general problem of system that interacts with
its environment is of great importance in many
fields of physics. The basic ingredients of this
interaction are illustrated in Fig.15.
On the one hand we have the effect of dissipation,
meaning that energy is lost by the "system" (Brownian particle)
and is absorbed by the "environment" (gas particles).
On the other hand the environment induces fluctuations
that acts like "noise" on the system.
The "noise" has two significant effects: One is to
pump "thermal" energy into the system, and the
other is to spoil quantum coherence. The latter
effect is called decoherence.

In case of bounded system, in the absence of
external time dependent fields, the interplay
between "noise" and "dissipation" leads eventually
to "thermalization". One may say that in
the thermal state the effect of dissipation
is balanced by the energy which is pumped by
the noise. Thus, both classically and quantum
mechanically we have to distinguish between
a "damping" scenario and an "equilibrium" situation.
The thermalization process is traditionally
described as "irreversible". On the other hand we
have the issue of "recurrences". We discuss the
latter issue in Section~20.

A systematic approach for the study of the dynamics
of a "system", taking into account the influence
of its environment, has been formulated
by Feynman and Vernon \cite{FV}.
The state of the system is represented by
the probability matrix $\rho(x',x'')$.
It is assumed that initially the "system" is prepared
is some arbitrary state. Its state at a later time
is obtained by a propagator $K(x',x''|x'_0,x''_0)$
which acts on the initial preparation.
The calculation of this propagator
involves a double path integral over all the
possible trajectories $x_A(t)$ and $x_B(t)$
that connect $(x'_0,x''_0)$ with $(x',x'')$.
This double path integral incorporates the effect
of the environment via an "influence functional"
which is defined as follows:
\begin{eqnarray} \label{e32}
F[x_A,x_B] \ \ = \ \
\langle \ U[x_B] \Psi_0 \ | \ U[x_A] \Psi_0 \ \rangle
\end{eqnarray}
Here $\Psi_0$ is the initial state of the environment.
If the environment is in "mixed" state, typically
a canonical (thermal) state, then the influence functional
should be averaged accordingly.

The absolute value of the influence functional can
be {\em re-interpreted} as arising from the interaction
with a \mbox{c-number} noisy field (with no back reaction).
The "phase" of the influence functional can be regarded
as representing the effect of "friction" (back reaction).
Thus there is one to one correspondence between
the Feynman-Vernon formalism, and the corresponding
classical Langevin approach. Note however that the
distinction between "noise" and "friction" is a matter
of "taste". Some people regard this distinction meaningless.

It should be realized that the calculation of the influence
functional for a given environment takes us back to the
more restricted problem of considering a "driven system".
The influence functional $F[x_A,x_B]$ is nothing but the
survival amplitude for a driving scheme that
involves "time reversal" (Eq.(\ref{e15})).

\section{Entanglement, decoherence and irreversibility}

The definition of decoherence is not a trivial
matter conceptually. There are several equivalent ways
to think about decoherence. The most "robust"
approach is to define decoherence as the irreversible
entanglement of the system with the environment:
Let us describe the the state of the system using
the probability matrix $\rho(x',x'')$. If the
system is prepared in pure state then
$\mbox{trace}(\rho^2)=1$. Due to the interaction
with the environment the system gets entangled
with the environment, and as a result we will
have $\mbox{trace}(\rho^2) \le 1$. If the "environment"
consists of only "one spin", then we expect
$\mbox{trace}(\rho^2)$ to have "ups" and "downs",
and from time to time to have $\mbox{trace}(\rho^2)\sim 1$.
In such case we cannot say that the entanglement
process is "irreversible". But if the environment
consists of many degrees of freedom, as in the case
of interaction with "bath", then the loss of "purity"
becomes irreversible, and we regard it as a "decoherence process".

To be more specific let us consider the prototype
example of interference in Aharonov-Bohm ring geometry.
The particle can go from the input lead to the output lead
by traveling via either arms of the ring.
This leads to interference, which can be tested
by measuring the dependence of the transmission
on the magnetic flux via the ring. Consider now
the situation where there is a spin degree of freedom
in one arm \cite{imry}. The particle can cause a spin flip
if it travels via this arm. In such case interference
is lost completely. However, this entanglement
process is completely reversible. We can undo
the entanglement simply by letting the particle
interact with the spin twice the time.
Therefore, according to our restrictive definition,
this is not a real decoherence process.

Consider now the situation where a particle
gets entangled with bath degrees of freedom.
If the bath is infinite, then the entanglement
process is irreversible, and therefore it
constitutes, according to our definition,
a decoherence process.

At first sight it seems that for having irreversibility
one needs "infinity". This point of view is
emphasized in Ref.\cite{buttiker}:
Irreversibility can be achieved by having the
infinity of the bath (infinitely many oscillators),
or of space (a lead that extends up to infinity).
In this set of lectures we emphasize a third possibility:
Having irreversibility due to the interaction with chaos.
Thus we do not need "infinity" in order to have "irreversibility".

\section{Interpretation of decoherence as a dephasing process}

"Dephasing" is used as a synonym for "decoherence"
whenever a semiclassical point of view is adopted.
Determining the dephasing (decoherence) time
$\tau_{\varphi}$ for a particle $(x,p)$
that interacts with an environmental
degrees of freedom $(Q,P)$ is a central theme in
quantum physics. In the absence of such interaction
the $x$ motion is coherent, and interference should be
taken into account. This means, from semiclassical
point of view, that at least two trajectories
$x(\tau)=x_{\tbox{A}}(\tau)$ and $x(\tau)=x_{\tbox{B}}(\tau)$ have
a leading contribution to the probability to travel,
say, from $x(0)$ to $x(t)$, as in the prototype example
of the two slit experiment.

In the semiclassical framework the probability to
travel from one point to some other point is given
by an expression that has the schematic form
\begin{eqnarray} \label{e33}
\sum_{A,B} F[x_A,x_B]
\exp\left(i\frac{S[x_A]-S[x_B]}{\hbar}\right)
\end{eqnarray}
where $S[x]$ is the classical action, and $F[x_A,x_B]$
is the influence functional. Each pair of trajectories
is a "stationary point" of the Feynman-Vernon double
path integral. The "diagonal terms" are the so-called
classical contribution, while the "off-diagonal terms"
are the interference contribution.
It should be kept is mind that the validity of the
semiclassical approach is a subtle issue \cite{dph}.

The off-diagonal interference contribution is suppressed
due to the interaction with the environment if $|F[x_A,x_B]|\ll 1$.
Therefore $|F[x_A,x_B]|$ is called the "dephasing factor".
From the definition of the influence functional it is clear
that it reflects the probability to "leave a trace" in the
environment. Having $|F[x_A,x_B]|=0$ means that a different "trace"
is left in the environment, depending on whether the particle
goes via the trajectory $x_A(t)$ or via the trajectory $x_B(t)$.
In such case one can regard the interaction with the environment
as a "measurement" process.
In case of the DLD model (see Section~15)
this "trace" can be further interpreted as leaving an excitation
along the way. For critical discussion of this point
see Appendix C of \cite{qbm}. In the more general case
the notion of "leaving a trace" does not have a simple meaning.
All we can say is that decoherence means that the environment
is left in different (orthogonal) states depending on the
trajectory that is taken by the particle.

The law of "action and reaction" holds also in the world
of decoherence studies. Feynman and Vernon have realized that
the dephasing factor can be re-interpreted as representing
the effect of a c-number noise source (see section~16).
From this point of view the decoherence is due to the "scrambling"
of the relative phase by this noise. Hence the reason for using
the term "dephasing" as a synonym for "decoherence".
The analysis of dephasing using this latter point of
view can be found in \cite{qbm}. See also \cite{alicki}.
At high temperatures it is possible to use
a Markovian master equation approach (dynamical semigroups)
in order to obtain the (reduced) evolution of
the Brownian particle. The Markovian master equation approach
is described in other lectures of this school.
The master equation in case of the DLD model
can be found in Section~3 of \cite{qbm}.
Similar, but not identical master equations
are obtained in case of interaction with
many body bath \cite{vacchini}.

\section{Determination of the dephasing time}

In the above described semiclassical framework,
the problem of dephasing reduces to the more
restricted problem of studying the dynamics of a
time dependent Hamiltonian ${\cal H}(Q,P;x(t))$.
Moreover, we see that the Feynman-Vernon dephasing
factor is just the absolute value of the fidelity
amplitude $F(t)$ that corresponds to Eq.(\ref{e15}).
Note however that here we use a more general notion
of fidelity: The restricted definition of
fidelity (Eq.(\ref{e17})) is formally obtained
if $x_{\tbox{A}}(\tau)$ and $x_{\tbox{B}}(\tau)$
are "rectangular pulses".

The dephasing time $\tau_{\varphi}$ is defined as
the time that it takes for $|F(t)|$ to drop significantly
from $|F(t)|\sim 1$ to some very small value $|F(t)|\ll 1$.
Let us concentrate on the Brownian motion model
of Fig.16a. If the motion of the Brownian particle
is characterized by a velocity $V$, then we have to
distinguish between the following possibilities:
Having very small "adiabatic" velocities;
Having intermediate velocities that allow LRT treatment;
And having non-perturbative velocities. In the latter
case a semiclassical picture can be justified.

The detailed analysis of the problem can be found
in \cite{wld}. Here we just quote the final results.
In the semiclassical regime
\begin{eqnarray} \label{e34}
\tau_{\varphi} \  = \ \tau_{\tbox{col}} \ = \
\frac{L_{\tbox{col}}}{v_{\tbox{E}}}
\end{eqnarray}
where $L_{\tbox{col}}$ is the mean free path
between collisions with the Brownian particle.
This is the expected naive result. It means that
one collision with the Brownian particle
is enough in order to "measure" its trajectory.
The other extreme case in having extremely small
adiabatic velocities. To the extend that we can
trust adiabaticity there is no dephasing at all:
The gas particle simply "renormalize" the bare
potential, which is in fact the Born-Oppenheimer
picture. Of course, if we take into account
corrections to the adiabatic picture, then
we get a finite dephasing time. In the LRT regime
of velocities we can estimate the dephasing time as
\begin{eqnarray} \label{e35}
\tau_{\varphi} \ \ \approx \ \
\left(\frac{L_{\tbox{col}}\lambda_{\tbox{E}}^2}
{v_{\tbox{E}}V^2}\right)^\frac{1}{3}
\end{eqnarray}
Both results have re-interpretation within
the framework of an effective DLD/ZCL model.
See \cite{wld} for details.

\section{Recurrences}

Consider ice-cube inside a cup of hot water.
After some time it melts and disappears. But if we
wait long enough (without time reversal)
we have some probability to see the ice-cube
re-emerging due to recurrences.
The issue of recurrences becomes relevant
whenever we consider a closed (un-driven) system.
In other words, whenever we do not try to control
its evolution from the outside.

There are recurrences both in classical and quantal
physics. In the latter case the tendency for recurrences
is stronger due to the quasi-periodic nature of the
dynamics. However, if the time scale for recurrences is
long enough with respect to other relevant time scales,
then we can practically ignore these recurrences.
Actually it is useful to regard these recurrences
as "fluctuations", and to take the standpoint that
our interest is only in some "average"
or "likely" scenario.

The thermalization process of the
particle-environment system is traditionally
described as "irreversible".
Indeed, if the bath is infinite, then also
the time for recurrences of the particle-bath
system becomes infinite.
On the other hand, if the bath is finite,
then we have to consider the recurrences of the
particle-bath system. These recurrences
can lead back to the initial un-entangled state.

In practice "recurrences" do not constitute
a danger for "irreversibility".
The time to get un-entangled
by recurrences is extremely large
(typically larger than the age of the universe).
Assuming a chaotic environment, and ignoring
issues of level statistics, the time scale
for recurrences is at least the Heisenberg time
(inverse of the mean level spacing)
of the combined particle-environment system.
It scales like $\hbar^{-(d+d_0)}$ where $d_0$
and $d$ are the number of degrees of freedom
of the particle and the environment respectively.

It goes without saying that the above issue
of recurrences becomes irrelevant if the
$x$ motion is treated classically.
There is however a twist to this
latter statement in the case where
the time variation of $x$ is strictly periodic.
This is due to dynamical localization effect \cite{qkr}.
Note however that dynamical localization
is a very fragile effect: Even in case that
it is found, it turns out that it manifests
itself only after a time that scales
like $\hbar^{-(1+2d)}$, which is much larger
than the Heisenberg time of the environment \cite{rsp}.


\ \\

\noindent
{\bf Acknowledgments:}

Essential for the promotion of this line of study are
the collaborations with Tsampikos Kottos (MPI Gottingen),
and with Alex Barnett (Harvard), and lately with Diego
Wisniacki (Comision Nacional de Energia Atomica, Argentina).
I also thank Shmuel Fishman (Technion) for many useful
discussions, and Rick Heller (Harvard), Joe Imry (Weizmann),
Bilha Segev (Ben-Gurion) and Uzy Smilansky (Weizmann) for
interesting interaction.

\ \\



\clearpage


\begin{figure}[h]
\centerline{\epsfig{figure=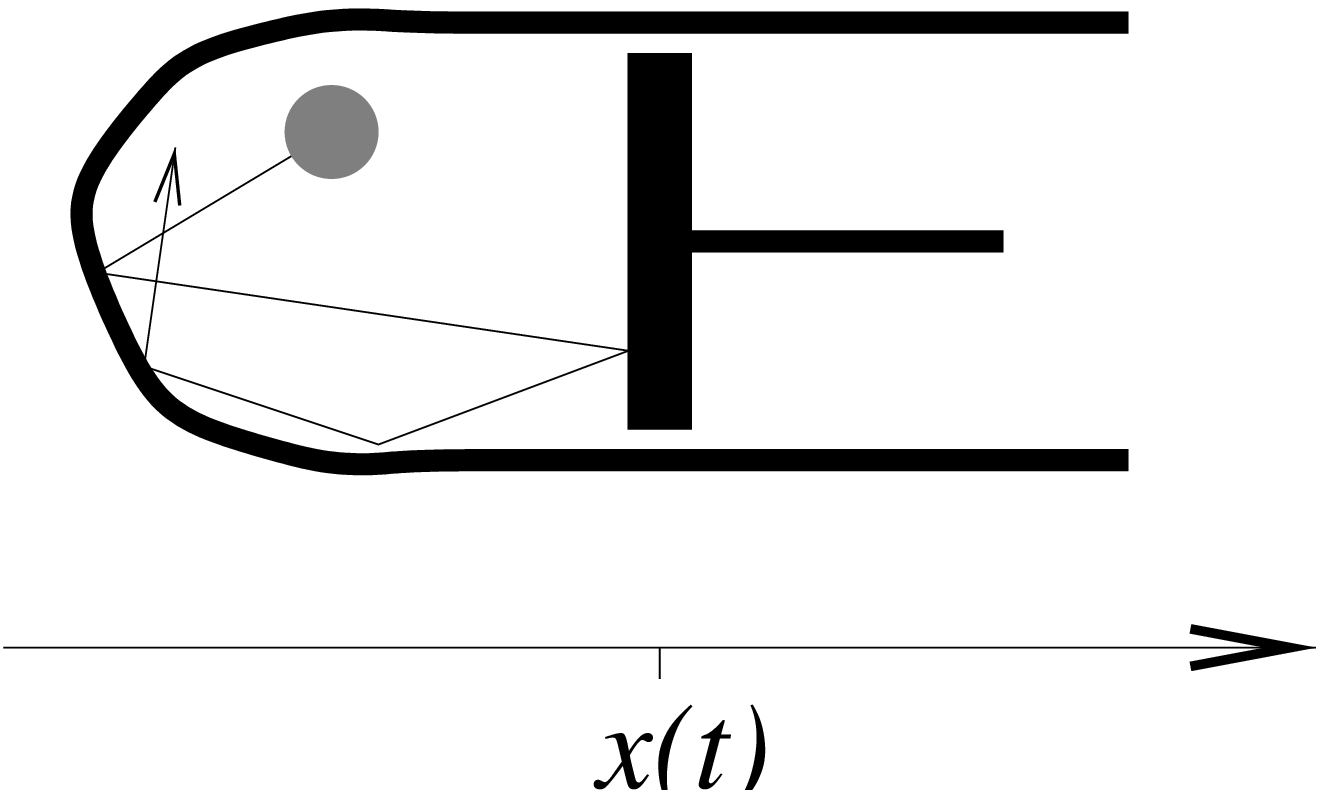,width=0.8\hsize}}
\caption{The prototype piston model. A gas particle
is moving chaotically inside a cylinder. The driving
is achieved by moving a wall element ("piston").}
\end{figure}

\vspace*{2cm}

\begin{figure}[h]
\centerline{\epsfig{figure=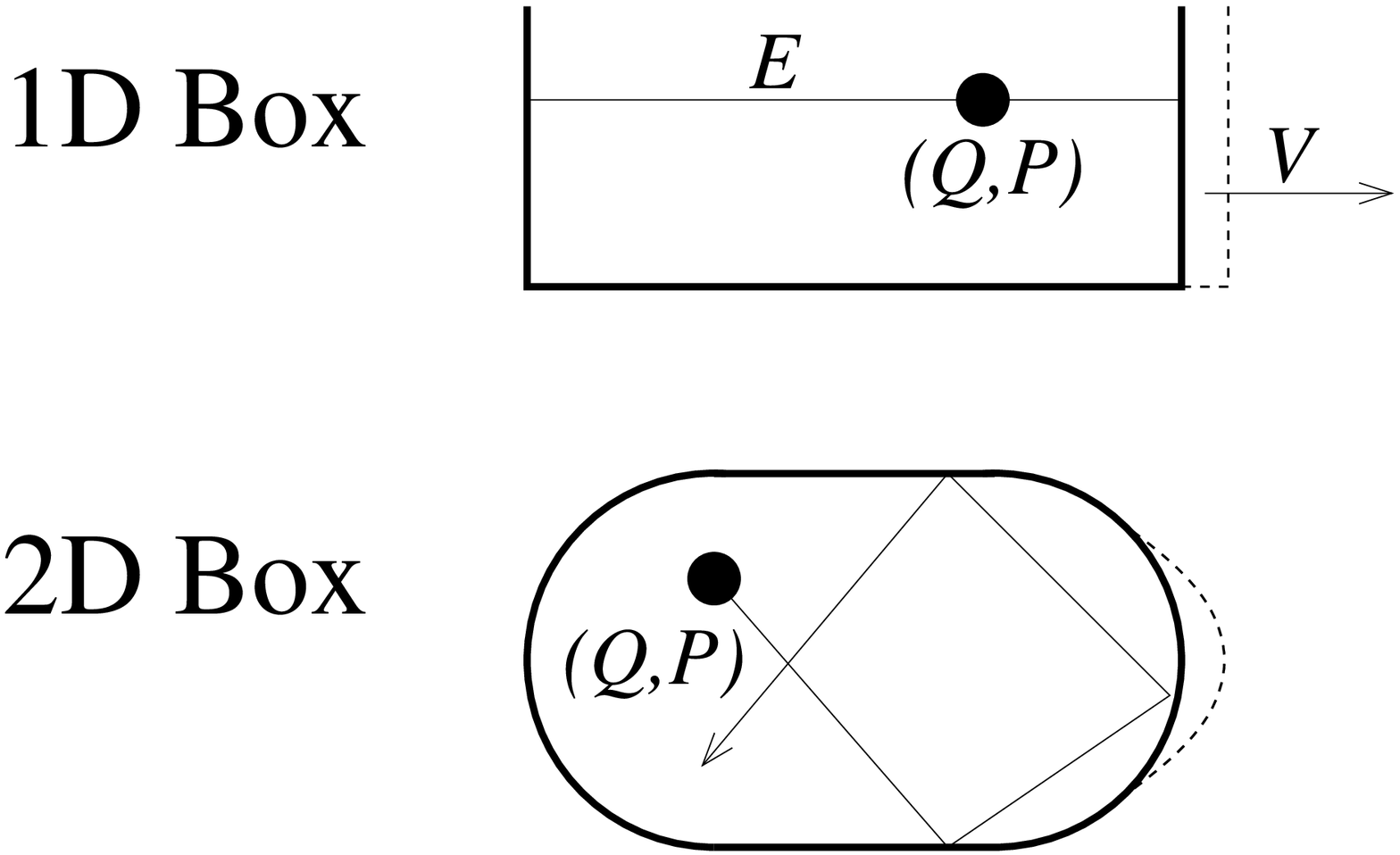,width=\hsize}}
\caption{The 1D version of the piston model (upper panel). 
The gas particle is moving inside an "infinite well". 
Its motion is not chaotic. In order to have chaotic 
motion we should consider at least a 2D box, for example 
a stadium shaped billiard system (lower panel). }
\end{figure}

\newpage

\begin{figure}[h]
\centerline{\epsfig{figure=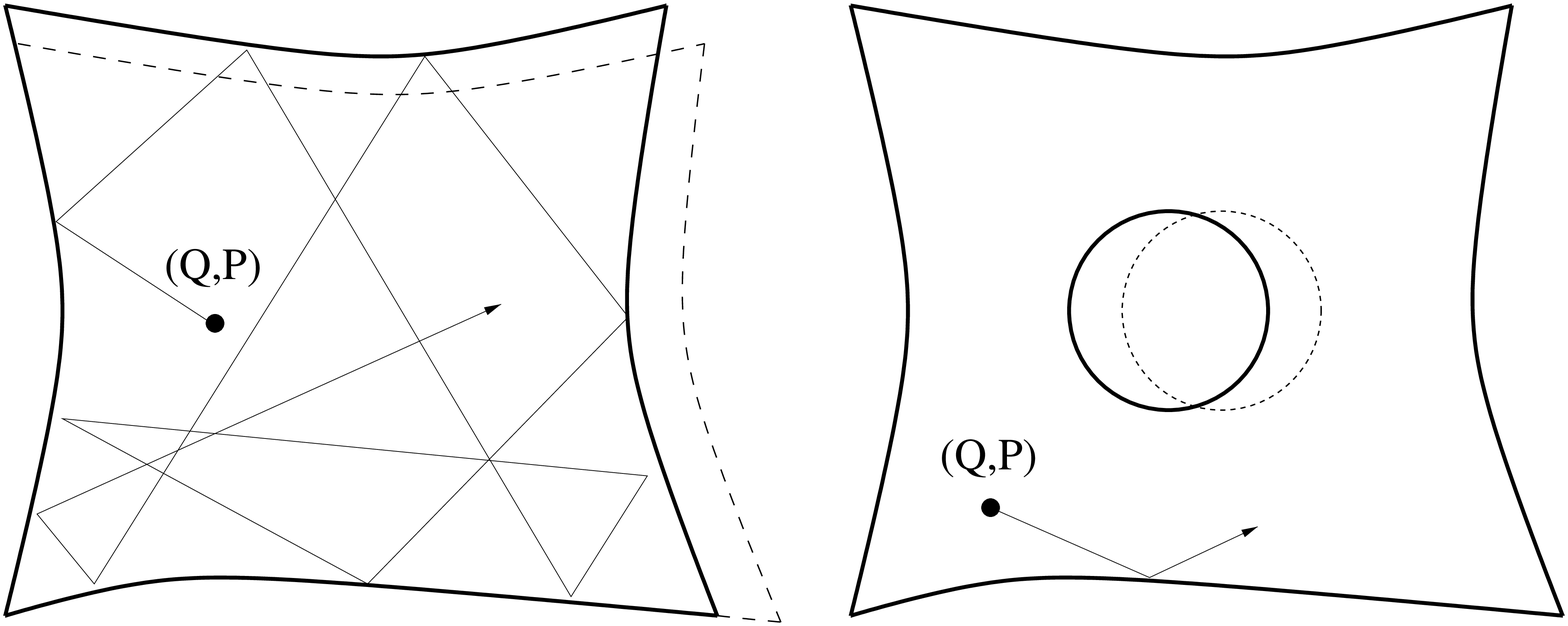,width=\hsize}}
\caption{Other versions of the "piston model". 
Two examples of "Sinai billiards" are illustrated.  
Note that in case of the right panel the displacement of 
the wall element manifestly does not involve a change 
of volume. This feature simplifies the analysis.}
\end{figure}

\vspace*{1cm}

\begin{figure}[h]
\centerline{\epsfig{figure=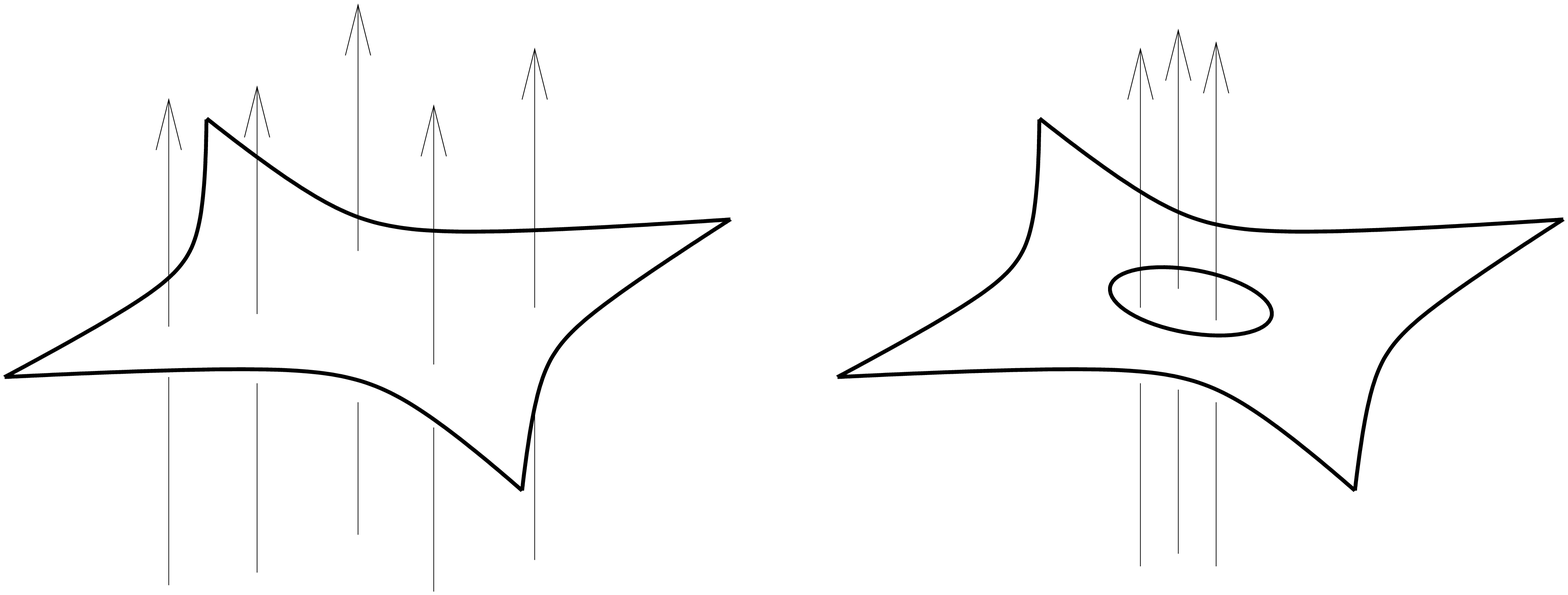,width=\hsize}}
\caption{The same "Sinai billiards" as in the previous
figure. Here the driving is achieved by changing the
perpendicular magnetic field. In case of the left panel,
where the box has a simple "quantum dot" geometry, the
magnetic field is assumed to be homogeneous.
In case of the right panel, where the box has
a aharonov-Bohm ring topology, the magnetic flux
is assumed to be concentrated in the hole.}
\end{figure}


\vspace*{1cm}

\begin{figure}[h]
\centerline{\epsfig{figure=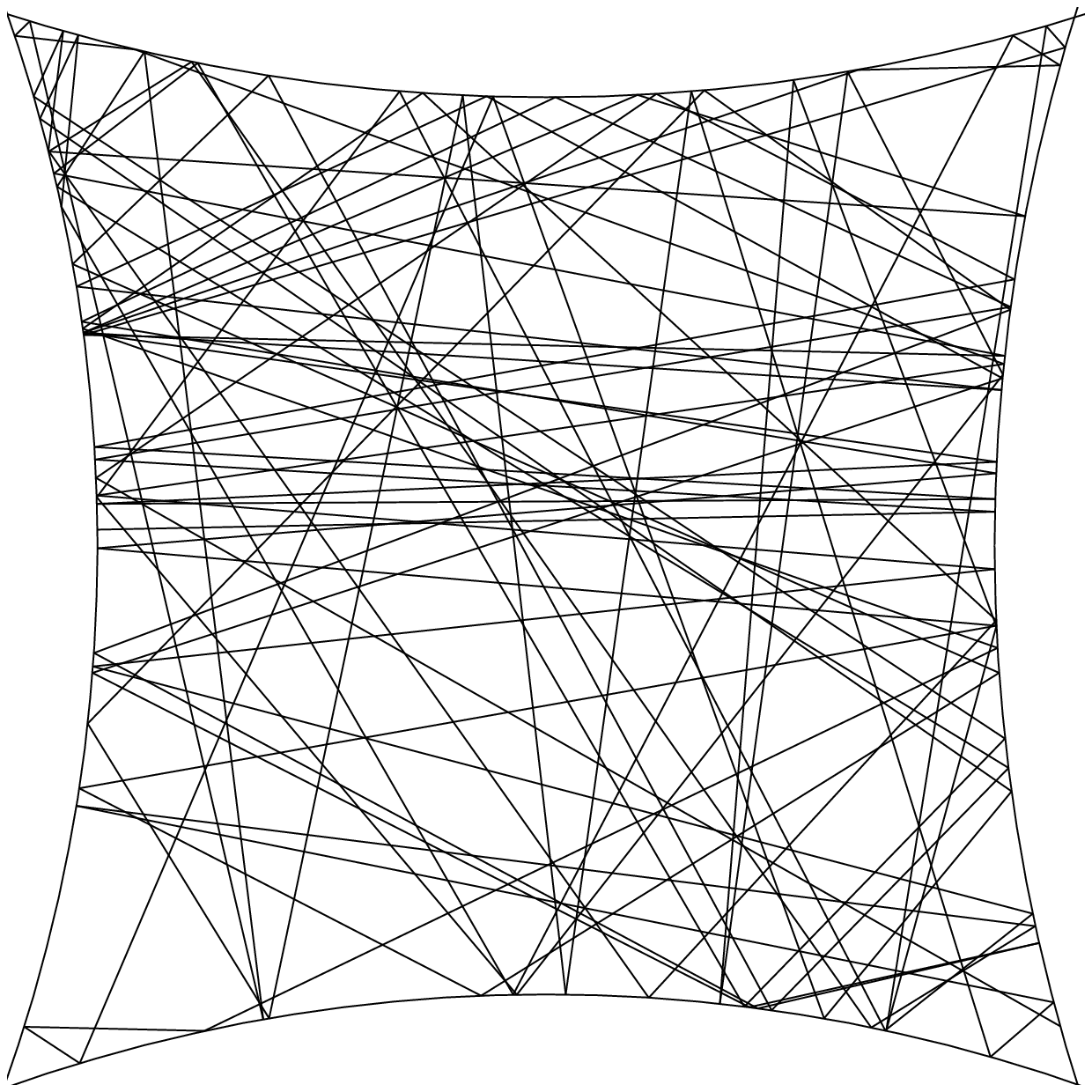,width=0.5\hsize}
\epsfig{figure=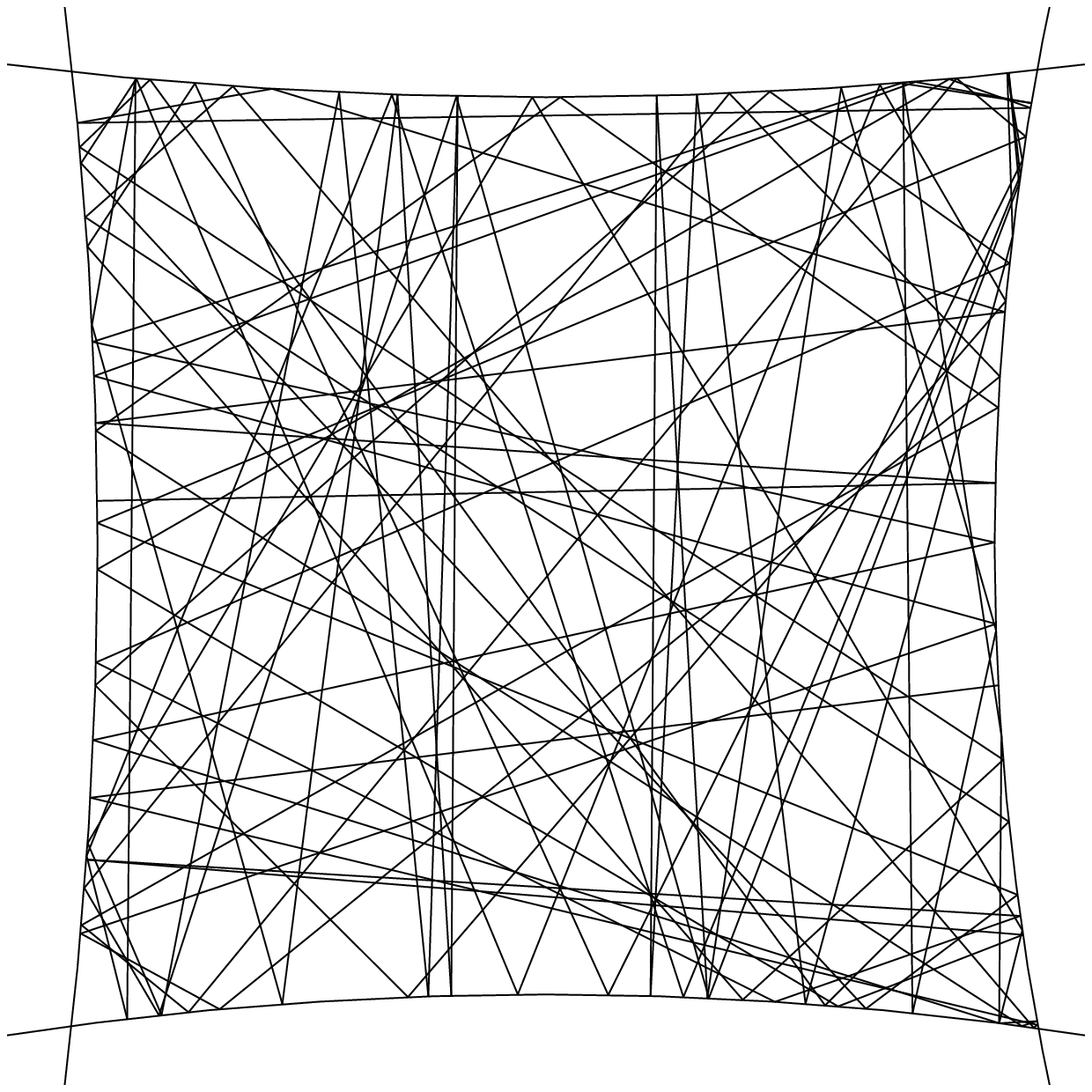,width=0.5\hsize}}
\caption{Chaotic trajectories inside a Sinai billiard.
In both examples the motion is completely chaotic.
This mean exponential sensitivity to any small change
in the initial conditions. This sensitivity can be
characterized by the "Lyapunov exponent". In the right
illustration the motion is chaotic, but the chaos is
weaker, which means smaller Lyapunov exponent.}
\end{figure}

\newpage

\begin{figure}[h]
\centerline{\epsfig{figure=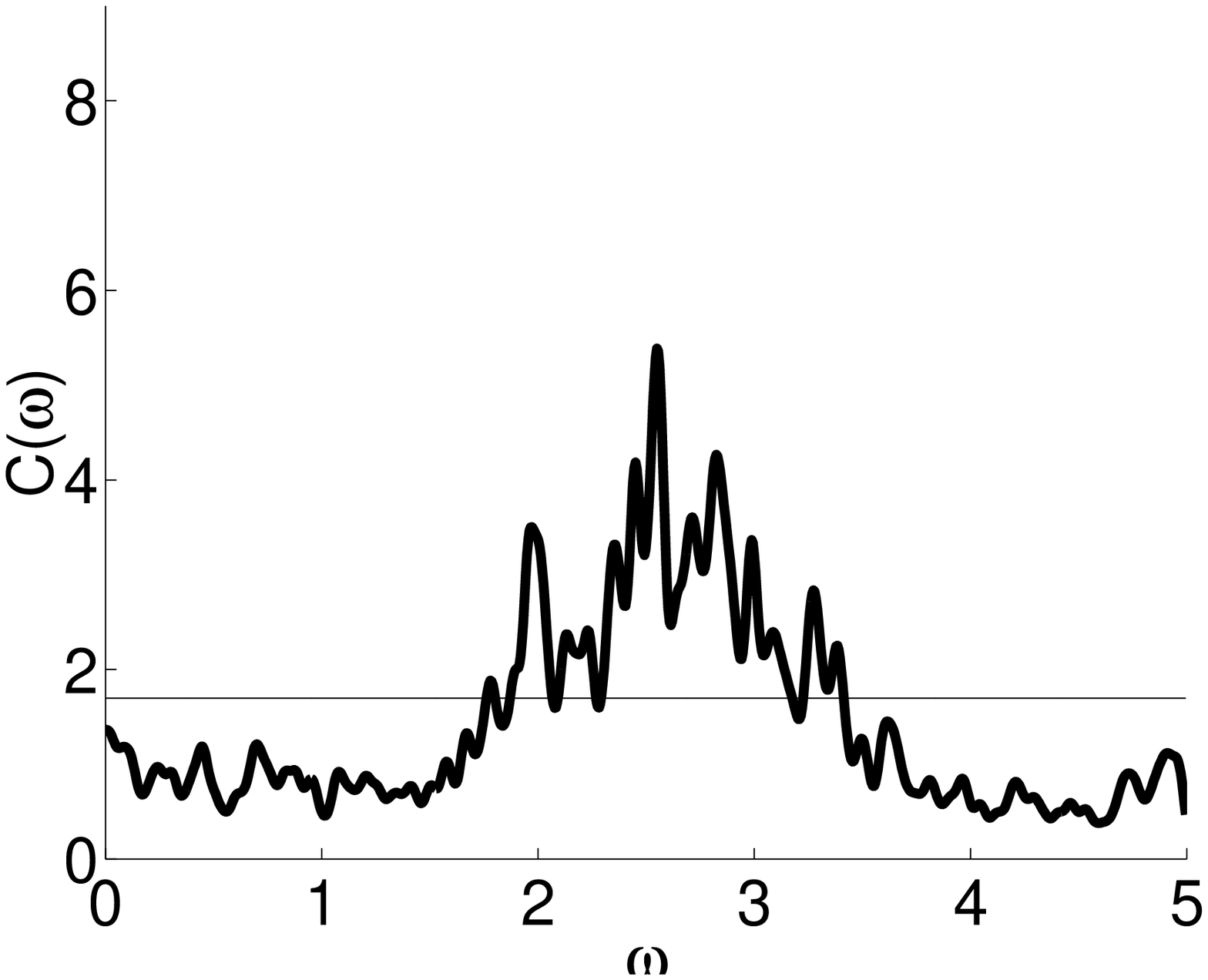,width=0.5\hsize}
\epsfig{figure=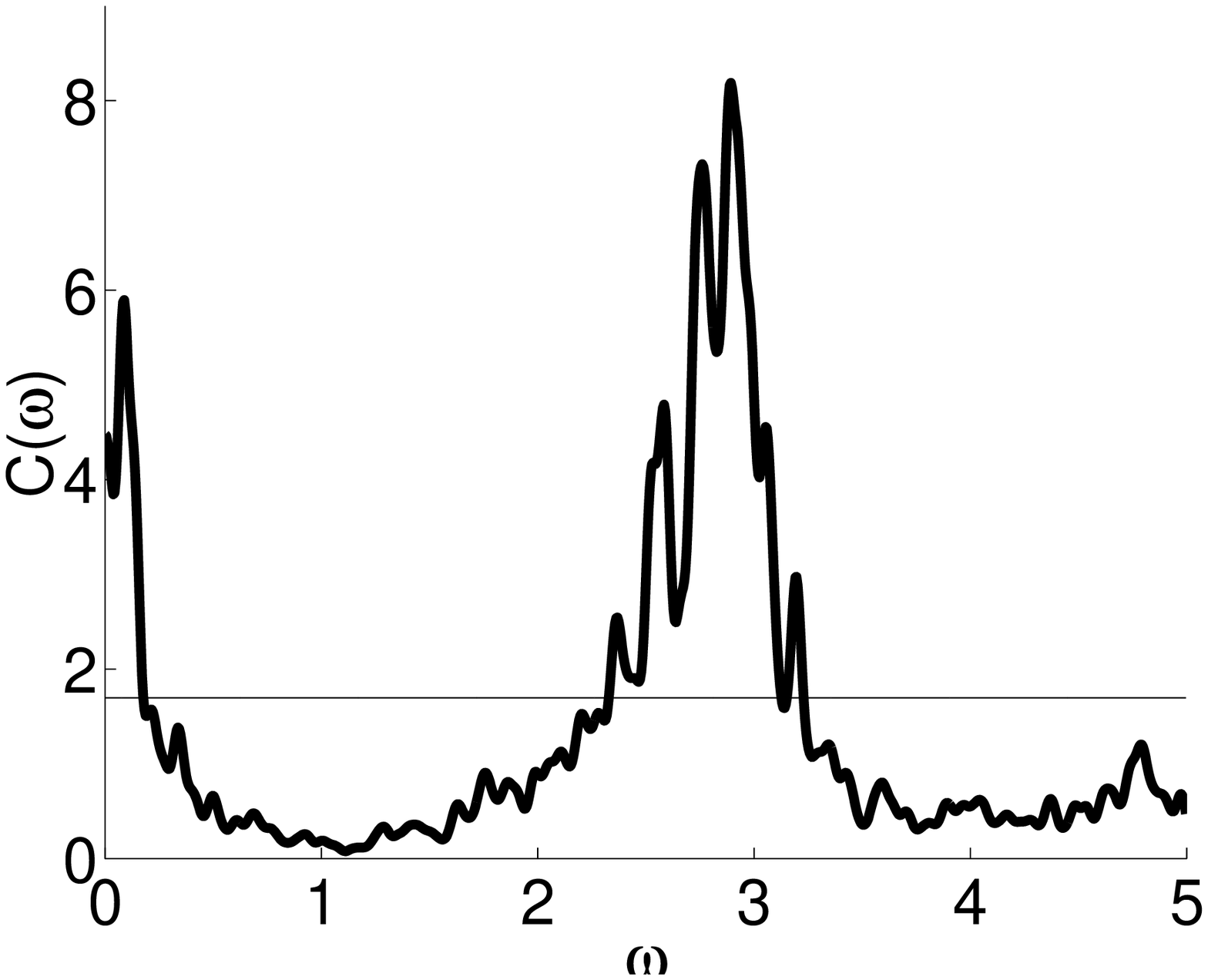,width=0.5\hsize}}
\caption{
The power spectrum of the motion for the two
examples of the previous figure.
It is the power spectrum $\tilde{C}(\omega)$
of the fluctuating quantity ${\cal F}(t)=-d{\cal H}/dx$.
The latter can be described as train of
impulses (spikes) due to collisions with the walls.
For strongly chaotic motion (left panel)
the power spectrum of ${\cal F}(t)$ is like that
of white noise.
In the right panel the bouncing frequency is
quite pronounced, and there is also a "diabatic" peak
around $\omega =0$. In both cases, the motion is
characterized by a continuous power spectrum, which
constitutes an indication for the chaotic nature of the motion.}
\end{figure}

\vspace*{2cm}

\begin{figure}[h]
\centerline{\epsfig{figure=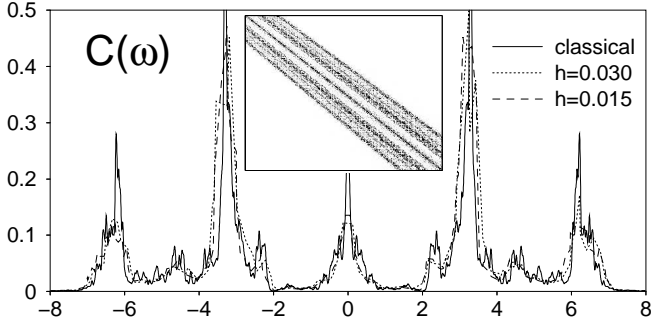,width=\hsize}}
\caption{
The band profile $(2\pi\hbar/\Delta)\cdot|\mbf{B}_{nm}|^2$
versus $\omega = (E_n{-}E_m)/\hbar$ is compared
with $\tilde{C}(\omega)$. See text for further explanations.
The calculation is done for the 2DW model of Eq.(\ref{e2}).
The inset is an image of a piece of the $\mbf{B}$ matrix.
Taken from Ref.\cite{lds}.}
\end{figure}

\newpage

\begin{figure}[h]
\centerline{\epsfig{figure=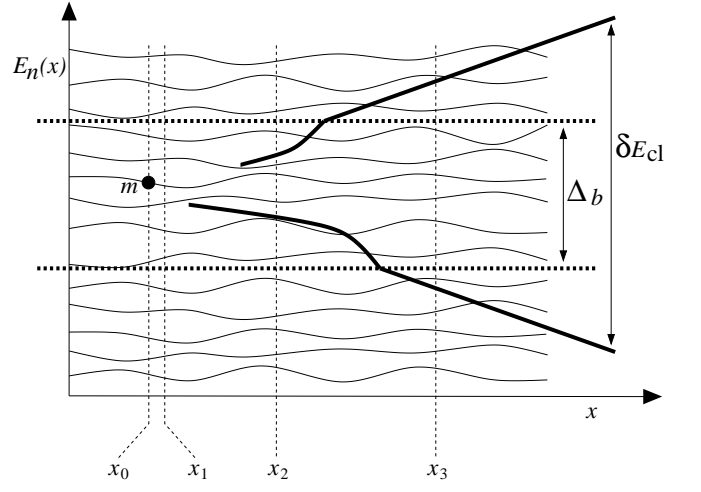,width=\hsize}}
\vspace*{2cm}
\centerline{\epsfig{figure=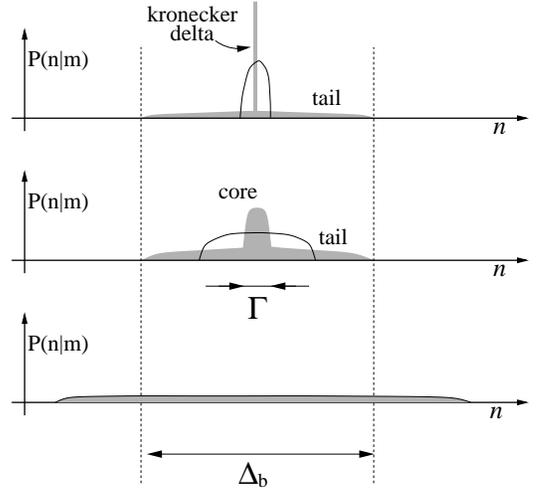,width=0.8\hsize}}
\caption{Upper panel: Schematic illustrations of the set of
energies $E_n(x)$ which are obtained via diagonalization
of a parameter dependent Hamiltonian.
The thick solid line indicates the $n$-range where 50\%
of the $P(n|m)$ probability is concentrated ($m$ is fixed).
The representative values $x_1$, $x_2$ and $x_3$ correspond
to the standard perturbative regime,
the core-tail (extended perturbative) regime,
and the non-perturbative regime respectively.
The corresponding LDOS structures are illustrated
(grey shading) in the three plots of the lower panel.
The semiclassical approximation (lines) is presented
for sake of comparison.}
\end{figure}


\newpage

\begin{figure}[h]
\centerline{\epsfig{figure=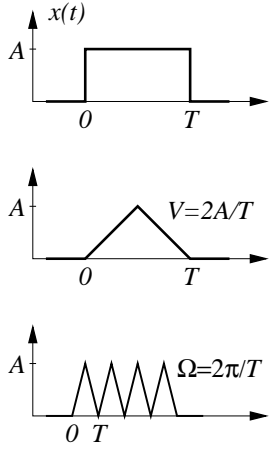,width=0.4\hsize}}
\caption{Various driving schemes:
Rectangular pulse of duration $T$ and amplitude $A$;
Triangular pulse which is further characterized
by finite driving "velocity" $V=|\dot{x}|=2A/T$;
Periodic driving with frequency $\Omega=2\pi/T$.}
\end{figure}

\vspace*{2cm}

\begin{figure}[h]
\centerline{\epsfig{figure=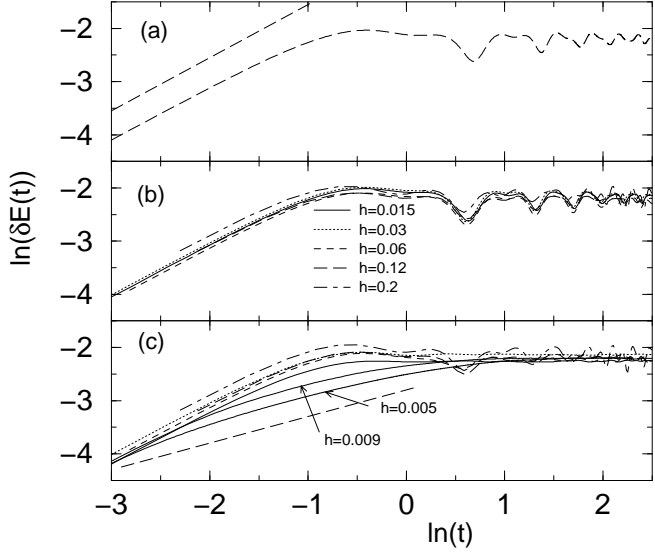,width=\hsize}}
\caption{
Energy spreading as a function of time for the 2DW
model: (a) classical; (b) quantum mechanical;
(c) An effective Wigner model. The energy in these
simulations is $E\sim 3$, and $\delta x = 0.2123$.
The light dashed lines in (a) and (c), that have slopes
$1$ and $1/2$ respectively, are drawn to guide the eye.
In (c) different lines correspond to different values
of $\hbar$ as in (b), and additional curves
($\hbar=0.009,0.005$) have been added.
Taken from Ref.\cite{wpk}.}
\end{figure}

\newpage

\begin{figure}[h]
\hspace*{0.07\hsize}\epsfig{figure=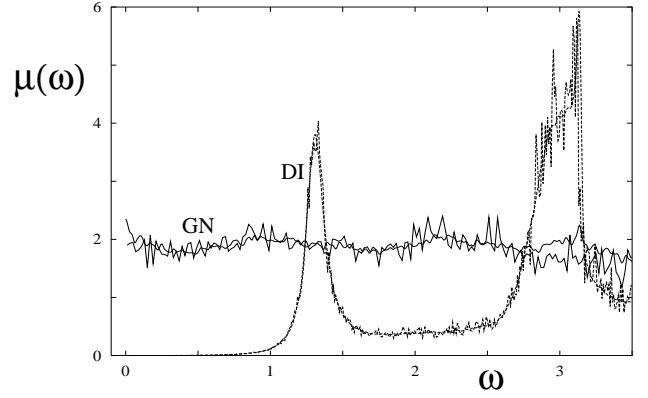,width=0.93\hsize}
\caption{The dependence of the friction coefficient
on the driving frequency, using LRT. "GN" stands for generic
deformation of a stadium shaped billiard, while "DI" stands
for special deformation (dilation). [In the latter case the
friction coefficient vanishes in the low frequency limit.]
In both cases the agreement between the classical (solid line)
and the quantum-mechanical (dashed line) calculation
is remarkable. Taken from Ref.\cite{dil}.}
\end{figure}

\vspace*{2cm}

\begin{figure}[h]
\hspace*{0.07\hsize}\epsfig{figure=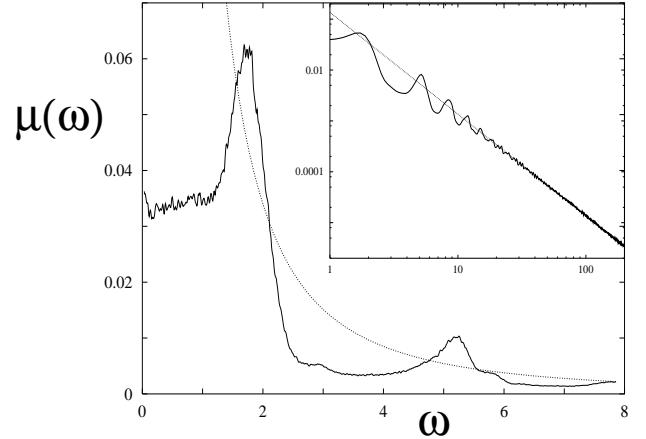,width=0.93\hsize}
\caption{The dependence of the mesoscopic conductance
on the driving frequency. The calculation is done
for a Sinai billiard shaped quantum dot, using LRT.
The result can be regarded  as a mesoscopic version
of Drude formula. The specific geometry of the system
is reflected in the structure of the response curve.
The inset is log-log plot. Taken from Ref.\cite{wlf}.}
\end{figure}

\clearpage

\begin{figure}[h]
\centerline{\epsfig{figure=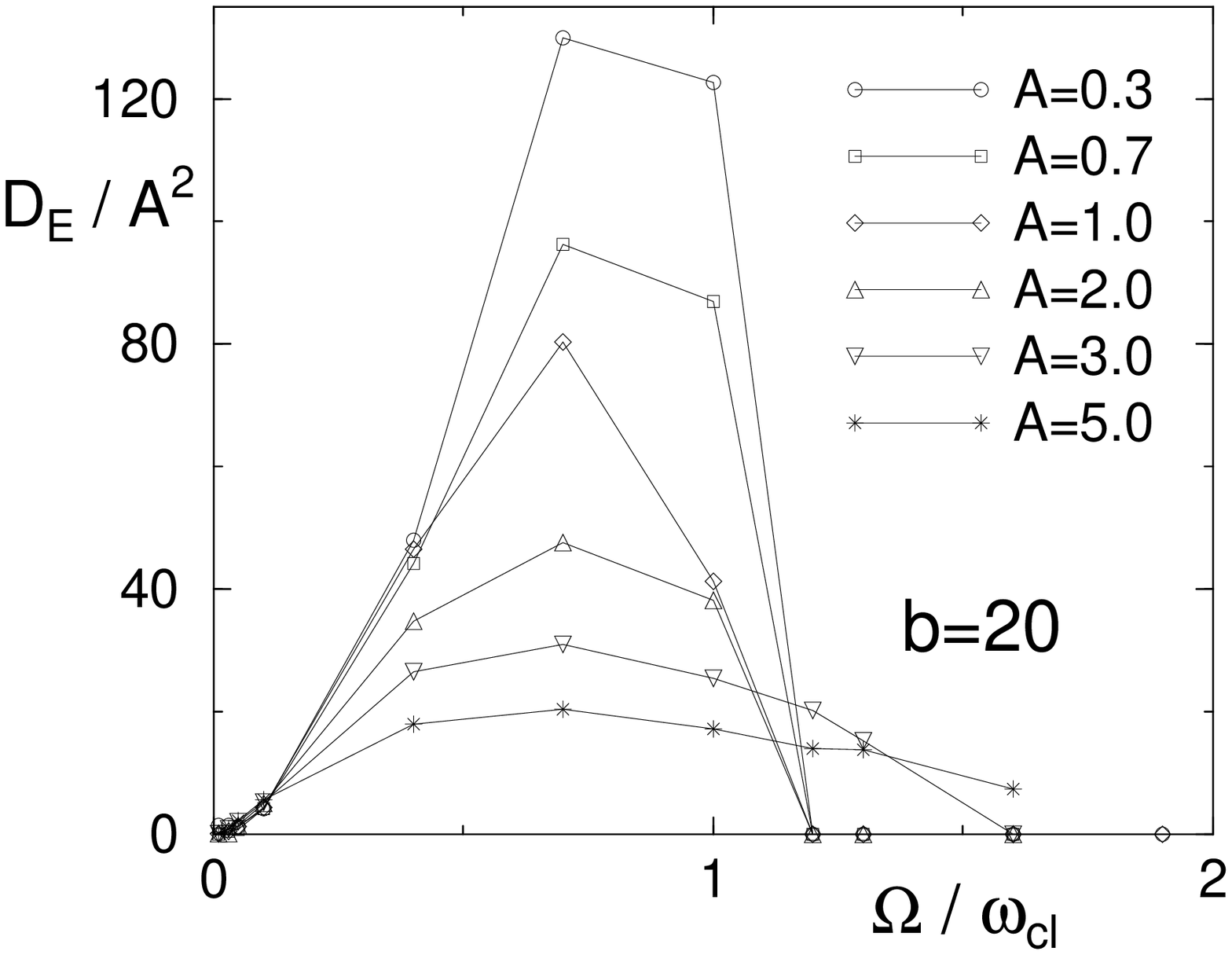,width=\hsize}}
\vspace*{1.5cm}
\centerline{\epsfig{figure=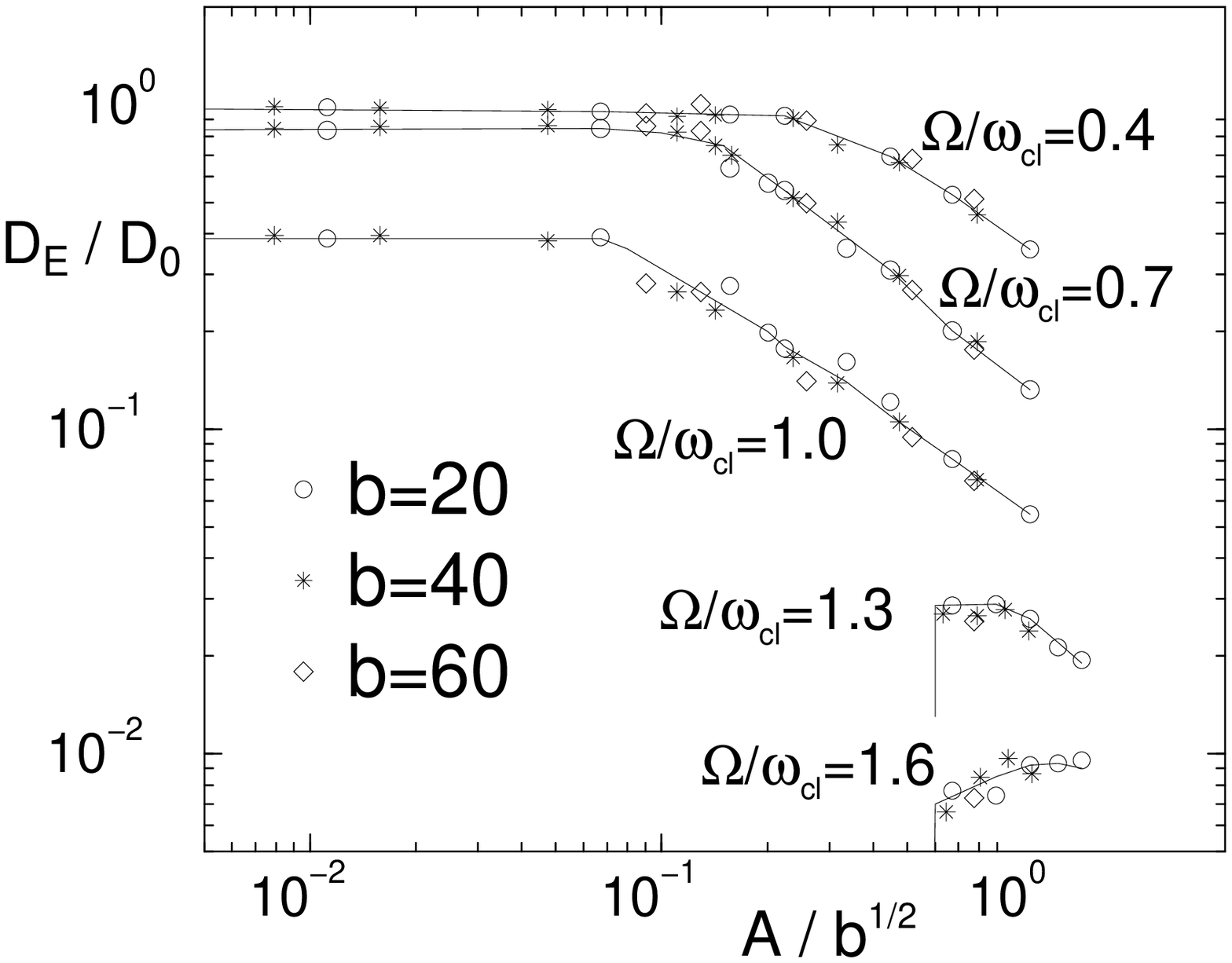,width=\hsize}}
\caption{
The response of a quantum mechanical system is
displayed as a function of $A$ and $\Omega$.
The simulations are done for Wigner model with
periodic driving. The units are chosen such
that $\Delta=0.5$ and $\hbar=1$ and $\sigma=1$.
{\bf Upper panel:}
Plots of $D_{\tbox{E}}/A^2$ versus $\Omega/\omega_{\tbox{cl}}$
for few values of $A$. For small $\omega$ the plots
coincide as expected from LRT. As $A$ becomes larger the
deviations from LRT scaling become more pronounced,
and we get response also for $\Omega>\omega_{\tbox{cl}}$.
{\bf Lower panel:}
Plots of $D_{\tbox{E}}/D_0$ versus $A/\sqrt{b}$
for few values of $\Omega/\omega_{\tbox{cl}}$.
LRT implies $D_{\tbox{E}}/D_0=1$ for $\Omega/\omega_{\tbox{cl}}<1$
and $D_{\tbox{E}}/D_0=0$ for $\Omega/\omega_{\tbox{cl}}>1$.
The purpose of the horizontal scaling is to demonstrate
that $A_{\tbox{prt}}$ rather than $A_c$ is responsible for
the deviation from this LRT expectation.
Taken from Ref.\cite{rsp}.}
\end{figure}

\newpage

\begin{figure}[h]
\centerline{\epsfig{figure=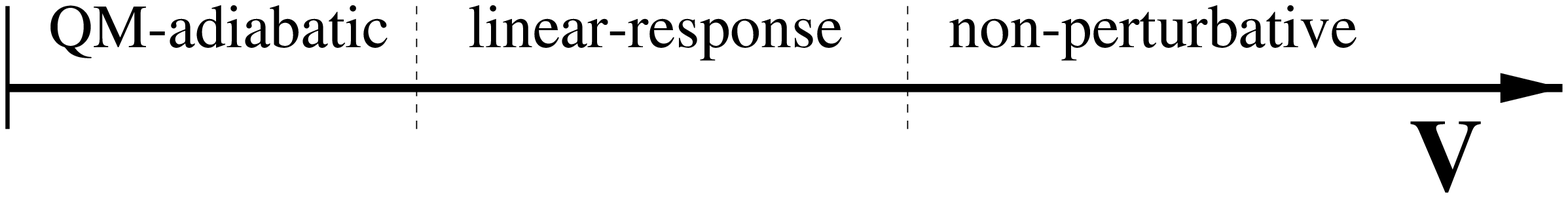,width=\hsize}}
\vspace*{2cm}
\centerline{\epsfig{figure=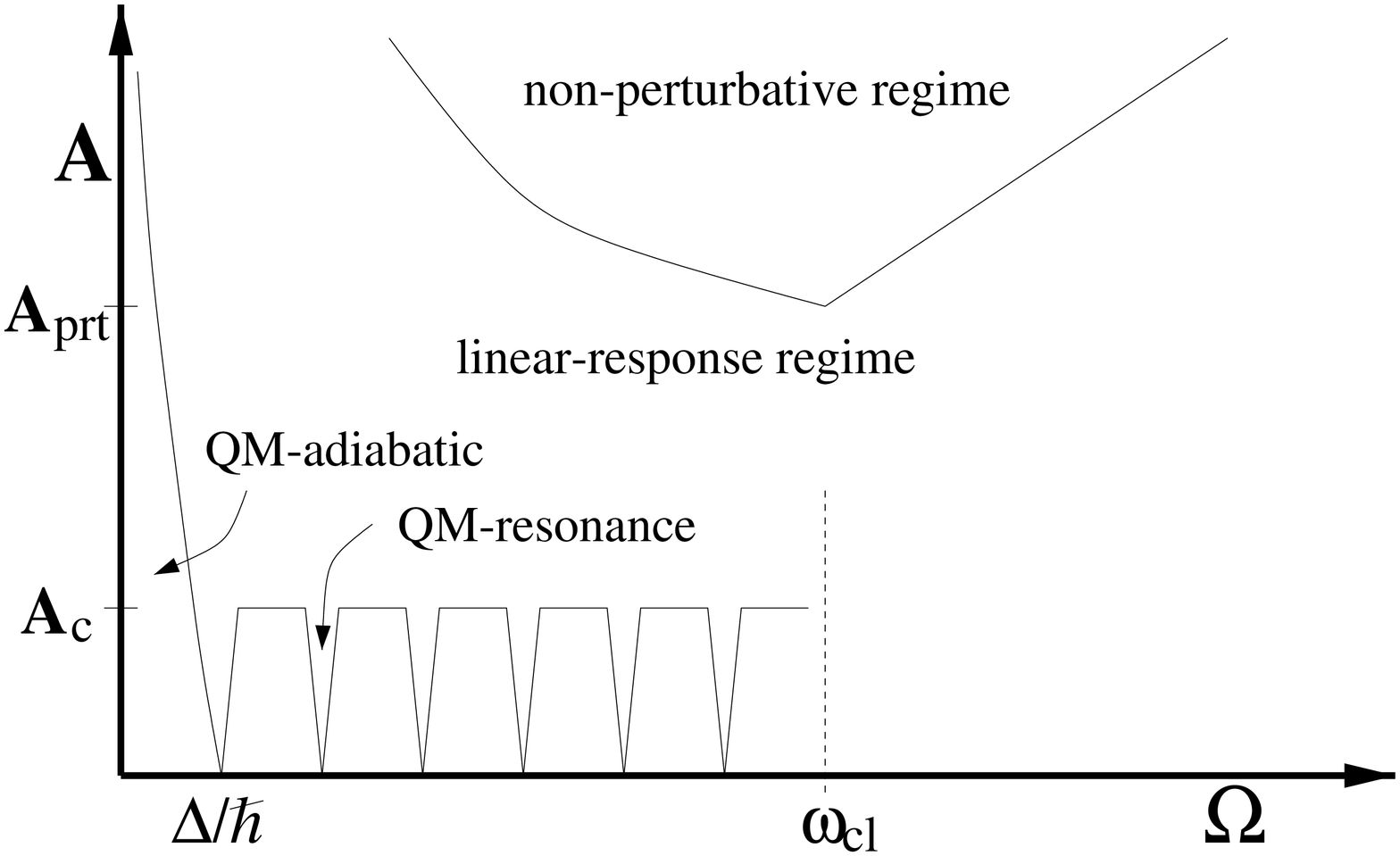,width=\hsize}}
\caption{
Upper diagram: The various $V$ regimes in the theory
of driven systems for linear driving $x(t)=Vt$.
Lower diagram: The various $(\Omega,A)$ regimes
for periodic driving $x(t)=A\sin(\Omega t)$.
We use the notations $\omega_{cl}=2\pi/\tau_{cl}$
and  $A_c=\delta x_c$
and $A_{\tbox{prt}} = \delta x_{\tbox{prt}}$.}
\end{figure}

\vspace*{2cm}

\begin{figure}[h]
\centerline{\epsfig{figure=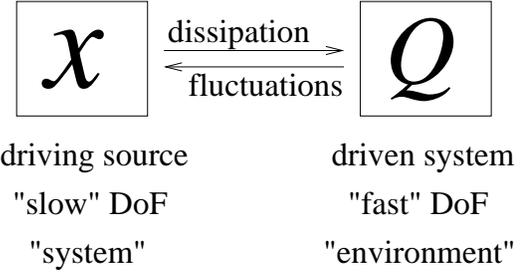,width=0.8\hsize}}
\caption{Block diagram illustrating the interaction
between system ($x$) and environmental ($Q$) degrees
of freedom (DoF). See discussion in the text.}
\end{figure}


\clearpage

\begin{figure}[h]
\centerline{\epsfig{figure=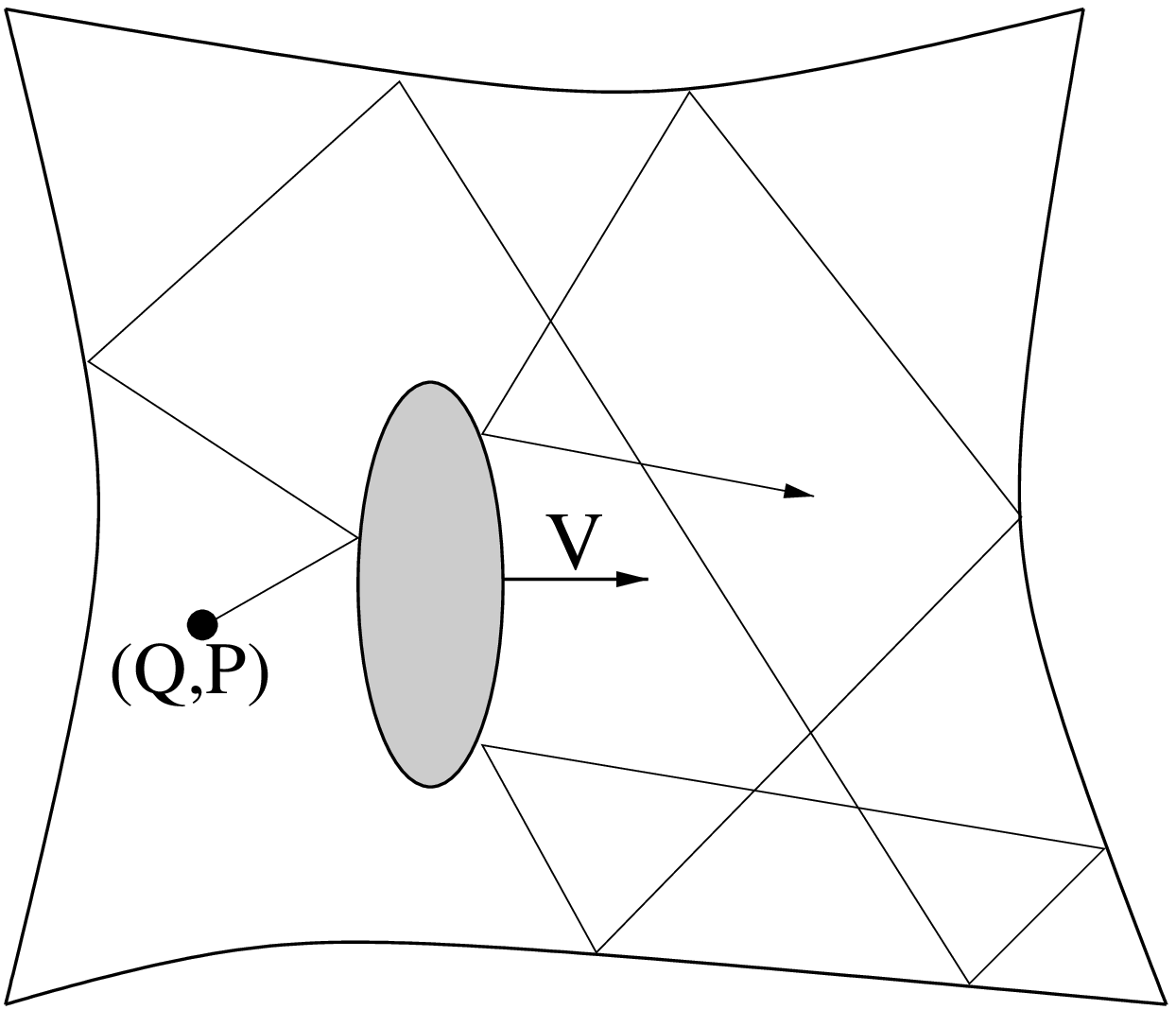,width=0.61\hsize}}
\vspace{0.6cm}
\centerline{\epsfig{figure=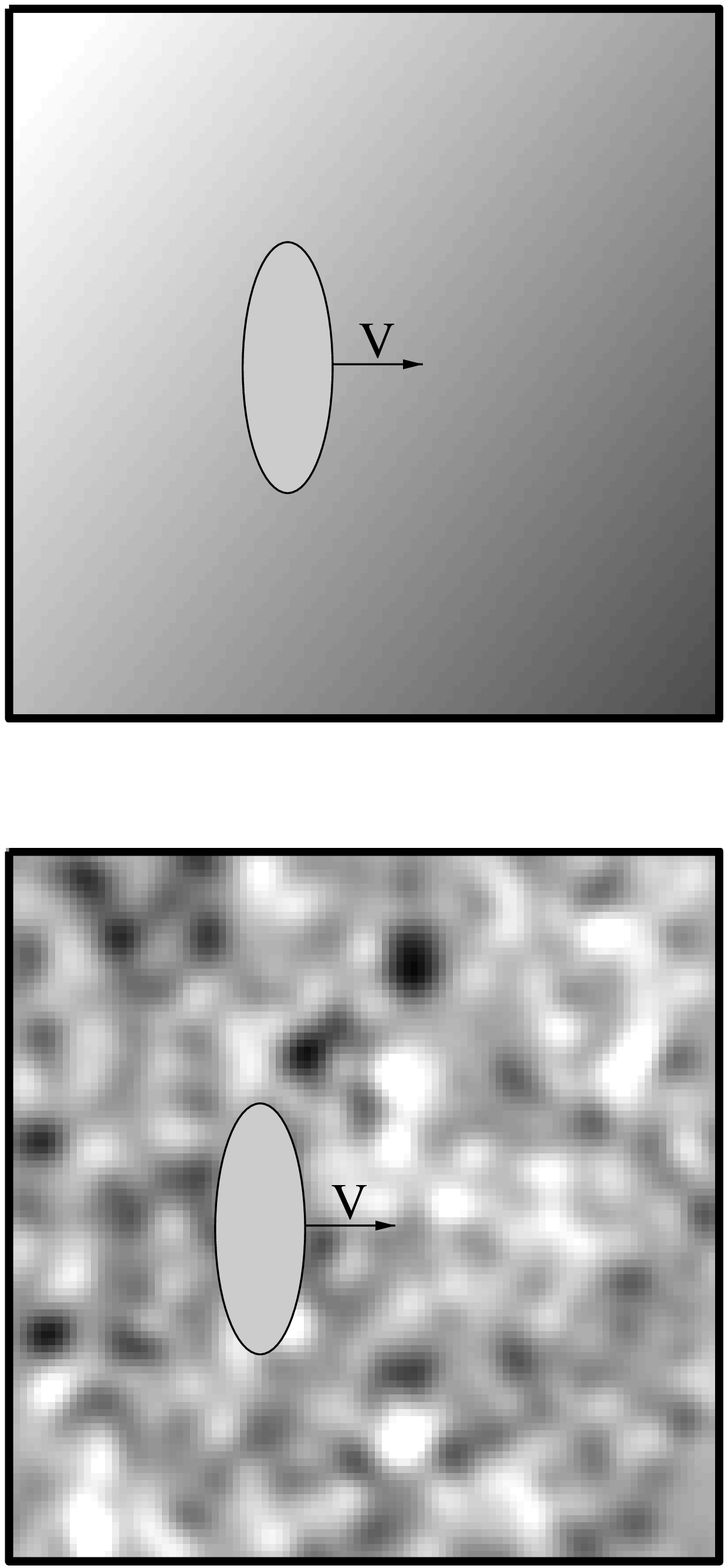,width=0.61\hsize}}
\vspace{0.5cm}
\caption{
(a) The Brownian motion is induced due to the
interaction with chaotic degrees of freedom.
(b) The Brownian particle in the ZCL model
experiences a fluctuating homogeneous field of force.
(c) In case of the DLD model the fluctuating field
is farther characterized by a finite correlation distance.
In (b) and (c) the background image is a
"snapshot" of the fluctuating environment.
Namely, the gray levels correspond to
the "height" of an instantaneous potential
which is experienced by the Brownian particle.}
\end{figure}

\newpage

\begin{figure}[h]
\centerline{\epsfig{figure=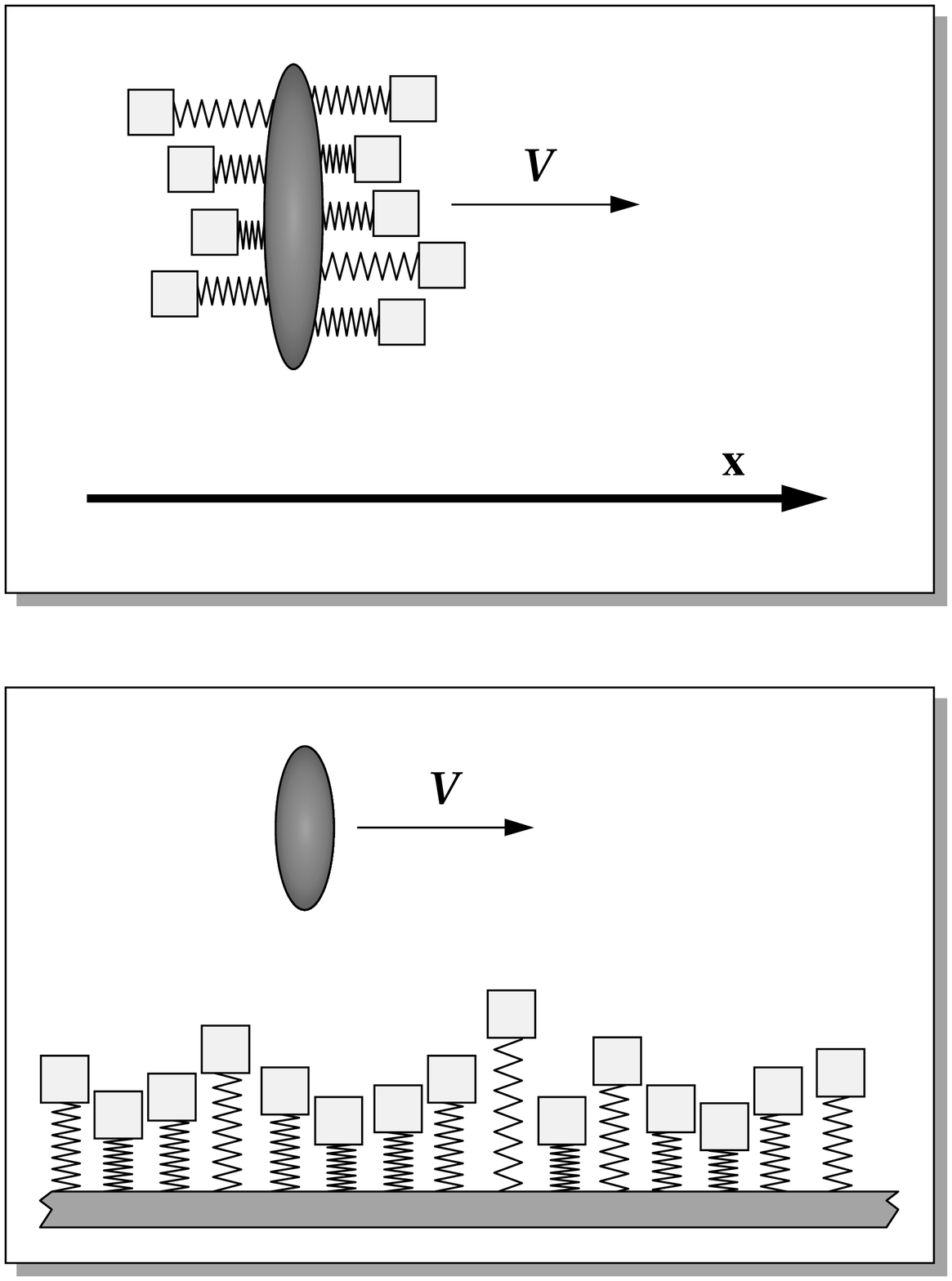,width=0.9\hsize}}
\caption{Schematic illustration of the ZCL model (upper panel)
and the DLD model (lower panel). The Hamiltonian of these
"spring systems" is literally the ZCL model and the DLD
model respectively. In the latter case the height of the masses
should be interpreted as the  "height" of the potential
which is experienced by the particle.}
\end{figure}

\begin{figure}[h]
\centerline{\epsfig{figure=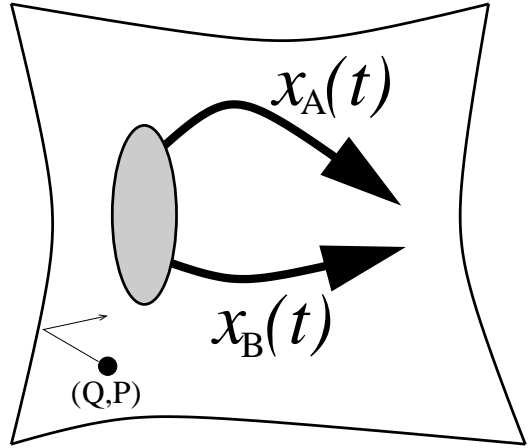,width=0.8\hsize}}
\caption{Schematic illustration of an interference experiment
using a semiclassical point of view. The Brownian particle
can take either the $x=x_A(t)$ trajectory, or
the $x=x_B(t)$ trajectory as in a two slit experiment.}
\end{figure}


\begin{thebibliography}{99}
\bibitem{cl_dsp}
See \cite{frc} and \cite{vrn}. The classical theory that is
presented in those references integrates ideas that were promoted in
previous studies, mainly \cite{ott} and \cite{wilk} and \cite{jar93}.
\bibitem{flc}
A clear formulation of the diffusion dissipation
relation can be found in \cite{vrn,frc},
and also in Appendix A of \cite{wlf}. It constitutes
a refinement/generalization of the dissipation
picture which is presented in \cite{wilk}.
The standard textbook formulation
of LRT and the fluctuation-dissipation relation
can be found in Appendix A of \cite{imry}.
\end{thebibliography}
\end{document}